\documentclass[preprint,prc,showpacs,showkeys,longbibliography]{revtex4-1}
\usepackage{epsfig,graphicx,float,color}
\usepackage{amssymb}
\usepackage[normalem]{ulem}

\begin{document}

\title{The quest of shape coexistence in Zr isotopes}

\author{J.E.~Garc\'{\i}a-Ramos$^{1,2}$ and K.~Heyde$^3$}
\affiliation{
$^1$Departamento de  Ciencias Integradas y Centro de Estudios Avanzados en F\'isica, Matem\'atica y Computaci\'on, Universidad de Huelva,
  21071 Huelva, Spain\\
$^2$Instituto Carlos I de F\'{\i}sica Te\'orica y Computacional,  Universidad de Granada, Fuentenueva s/n, 18071 Granada, Spain\\
$^3$Department of Physics and Astronomy, Ghent University, Proeftuinstraat, 86 B-9000 Gent, Belgium}
\begin{abstract} 
\begin{description}
\item [Background] The mass region with $A\approx 100$ and $Z\approx 40$ is known to experience a sudden onset of deformation. The presence of the subshell closure $Z=40$ makes feasible to create particle-hole excitations at a moderate excitation energy and, therefore, likely intruder states could be present in the low-lying spectrum. In other words, shape coexistence is expected to be a key ingredient to understand this mass region.  

\item [Purpose] The aim of this work is to describe excitation energies, transition rates, radii, and two-neutron separation energies for the even-even $^{94-110}$Zr nuclei and, moreover, to obtain information about wave functions and deformation.  
 
\item [Method] The interacting boson model with configuration mixing will be the framework to study the even-even Zr nuclei, considering only two types of configurations: 0particle-0hole and 2particle-2hole excitations. On one hand, the parameters appearing in the Hamiltonian and in the $E2$ transition operator are fixed trough a least-squares fit to the whole available experimental information. On the other hand, once the parameters have been fixed, the calculations allow to obtain a complete set of observables for the whole even-even Zr chain of isotopes.     

\item [Results] Spectra, transition rates, radii, $\rho^2(E0)$, and two-neutron separation energies have been calculated and a good agreement with the experimental information has been obtained. Moreover, a detailed study of the wave function has been conducted and mean-field energy surfaces and deformation have been computed too.
  
\item [Conclusions] The importance of shape coexistence  has been shown to correctly describe the $A\approx 100$ mass area for even-even Zr nuclei. This work confirmed the rather spherical nature of the ground state of $^{94-98}$Zr and its deformed nature for $^{100-110}$Zr isotopes. The sudden onset of deformation in $^{100}$Zr is owing to the rapid lowering of a deformed (intruder) configuration which is high-lying in lighter isotopes. 
\end{description}
\end{abstract}



\date{\today}
\maketitle

\section{Introduction}
\label{sec-intro}
In recent years, it has become clear that nuclei can exhibit various shapes changing from more spherical structures into the characteristic of strongly deformed  systems.  It is even more interesting that a given nucleus can exhibit the possibility of various shapes, and this as a function of the excitation energy available in the nucleus.  This idea was first brought forward already in the 50's in the $^{16}$O nucleus by \citet{morinaga56}.  It was clearly at odds with the standard expectation of what a doubly closed shell nucleus was supposed to exhibit as its low-lying excited state spectrum, in particular, the appearance of a low-lying 0$^+$ state which nature was not clear \textit{a priori}. It turned out to be not the only doubly-closed shell nucleus to exhibit such shape coexistence structure.  The nuclei $^{40}$Ca, $^{56}$Ni, but not only, have more recently indicated the appearance of bands pointing to strongly deformed structures \cite{Idegu01,Johan08,Caur07,Tsuno14}. The same, what was called shape coexistence, was observed in the heavy single-closed shell nuclei such as the Pb isotopes and the Sn nuclei, with unambiguous indications of experimental evidence for the appearance of deformed bands (see \cite{heyde11} for a review). The wealth of data have given rise to intensive studies moving away from stability at Radioactive Ion Beam facilities to carry out an extensive set of Coulomb excitation experiments (see \cite{kasia19} and references therein), complementing the information on energy spectra, with the more informative data on $B(E2)$ values, quadrupole moments, $\rho^2(E0)$ values, and S$_{2n}$ values. On the theoretical side, with the increasing computing capabilities, in particular for the lighter nuclei, large-scale shell-model calculations (LSSM) have been carried out  during the last decade.  In contrast with former studies,  much large model spaces are handled, encompassing a number of adjacent major harmonic oscillator shells. Consequently,  starting from a given two-body interaction,  the essential correlations that the interacting nucleon systems can develop have shown to be able to give rise to the appearance of shape coexistence (see also \cite{Poves18}, \cite{Dario18}, \cite{Tsunoda18co} for some very recent papers).  Starting from calculations within a laboratory frame basis, care has to be taken when discussing the issue of ``nuclear shapes and its evidence'' because is a non-observable. Keeping with the fact that there are the data that reign, it is the comparison of data with the theoretical results the key point, and not so much shape properties, which imply some model-dependence. Mean-field studies, using the self-consistent Hartree-Fock-Bogoliubov (HFB) method, have extensively addressed these nuclei, pointing out the presence of various shapes as a function of changing number of protons (or neutrons), covering the larger part of the nuclear mass region \cite{Bender03,Dela10,niksic11,nomura08,nomura10} 
 
The aim of this work is, on one hand, to present in a coherent way the present experimental situation of even-even Zr isotopes and the different theoretical approaches to treat them. On the other hand, it will be carried out a very detailed interacting boson model calculation using configuration mixing (IBM-CM in short) to describe the even-even Zr nuclei. The paper is organized as follows: in Section \ref{sec-exp}, we describe the experimental situation in the Zr isotopes whereas in Section \ref{sec-theo}, we present the various theoretical approaches that have been used in the literature to study the Zr nuclei.  In Section \ref{sec-ibm-cm}, we succinctly present the IBM-CM formalism as well as the fitting methodology used. We also discuss the main outcome of the calculations such as energy spectra, $B(E2)$ values, and its comparison with the available experimental data. In Section \ref{sec-other}, we discuss the results on the isotopic shifts, E0 transition rates and two-neutron separation energies, and Section \ref{sec-q-invariants} is devoted to the description of nuclear deformation properties of the Zr nuclei as derived from the study of the quadrupole shape invariants and from IBM-CM mean-field calculations. Finally, in Section \ref{sec-conclu}, both the main conclusions as well as an outlook for further studies are presented and suggestions are made for future experiments to be carried out.

\section{Experimental data in the even-even Zr nuclei }
\label{sec-exp}

\begin{figure}[hbt]
\centering
\includegraphics[width=0.90\linewidth]{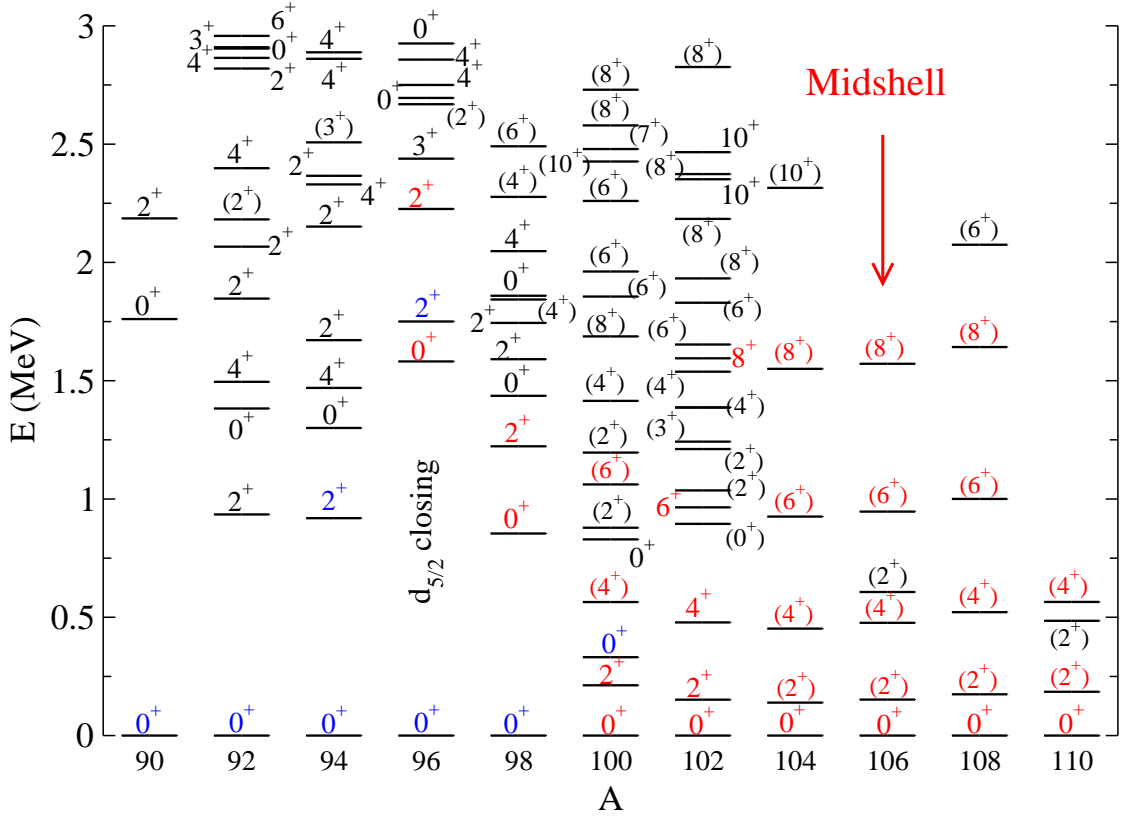}
\caption{Experimental energy level systematics for the Zr isotopes. Only levels up to $E_x \sim 3$ MeV are shown. Levels labelled in blue correspond to spherical shapes, while labelled in red to deformed ones (see text for details).}
\label{fig-e-systematics-zr}
\end{figure}

At present, data have become available for an extensive series of isotopes, spanning all the way from the closed subshell-closed shell nucleus $^{90}$Zr, passing through $^{100}$Zr up to $^{110}$Zr which was speculated about as a possible doubly subshell closing nucleus with $Z=40$ and $N=70$ within a spherical harmonic oscillator picture. The data go all the way down to the very early work by \citet{Cheif70} who at the LBL Lawrence Radiation Laboratory reported the first strong experimental evidence concerning a new region of deformation in the $A=100$ mass region and beyond. 

In Fig.~\ref{fig-e-systematics-zr},  we present the resulting data without making any connecting lines between the energy levels in order to avoid any bias at the start.  According to the literature, the levels labelled in blue corresponds to states with a clear spherical structure, while those in red to a deformed shape.  What is very clear though is a major change in the spectral characteristics. Once having passed neutron number $N=56$ (which corresponds to a filled 2d$_{5/2}$ orbital) and passing through the $N=98-100$ region, energy spectra clearly exhibit a more collective character in the ground-band structure. In the discussion following, besides information on the energy spectra, we emphasize the importance of both electromagnetic properties, isotopic shifts, S$_{2n}$ separation energies as well as on data for two-nucleon transfer reactions giving essential information to uncover the properties of this most interesting Zr set of isotopes, covering $\approx$ 20 mass units.

We have made use of the most recent Nuclear Data Sheets references, namely, \cite{Bagl12b} for $A=92$, \cite{Abri06} for $A=94$, \cite{Abri08} for $A=96$, \cite{Sing03} for $A=98$, \cite{Sing08} for $A=100$, \cite{Defr09} for $A=102$, \cite{Blac07} for $A=104$, \cite{Sing15} for $A=106$, and \cite{Sing15b} for $A=108$. Within the last $\sim 20$ years, extensive use has been made of a large set of complementary probes in order to characterize both the electromagnetic decay of the levels in the Zr series of isotopes, spanning the region from $A=90$ up to $A=110$.  This includes the measurement of nuclear lifetimes using Doppler-broadened lineshapes of the decaying $\gamma$ rays, making use of neutron scattering using the (n,n'$\gamma$) method as well as using the Doppler Shift Attenuation (DSA) technique as well as measurements of E0 $\rho^2$ values (see \cite{Wood99,Kibe05}).  More recently, Coulomb excitation methods have become state-of-the-art approach to extract not only the $B(E2)$ values but even the sign of specific $E2$ reduced matrix elements. Besides, the availability of high-resolution electron scattering allowed to measure $B(E2)$ values between the 0$^+_1$ and 2$^+_2$ states adding an independent approach to extract the $E2$ decay characteristics of excited 2$^+$. In order to reach the most heavy Zr isotopes, it was made use of prompt $\gamma$-rays in coincidence with isotopically identified fission products. In Table \ref{tab-references}, the most relevant experimental references are given.

\begin{table}[hbt]
  \caption{Relevant references concerning experimental data for even-even Zr isotopes.}
\label{tab-references}
\begin{center}
%  \begin{ruledtabular}
    \begin{tabular}{|c|l|}
      \hline
      Isotope & References\\
      \hline
      $^{90}$Zr &\citet{Warw79}, \citet{Garrett03}\\
      \hline
      $^{92}$Zr & \citet{Bagl12b}, \citet{Suga17}\\
      \hline
      $^{94}$Zr & \citet{Abri06}, \citet{Chak13}, \citet{Peters13},\\
               & \citet{Scheik14}, \citet{Elhalmi08}, \citet{Juli81} \\
      \hline
      $^{96}$Zr & \citet{Abri08}, \citet{Kumb03}, \citet{Alan16},\\
              & \citet{Krem16}, \citet{Piet18}, \citet{Kumb03}, \\
              & \citet{Sazo19}, \citet{Witt19} \\
      \hline    
      $^{98}$Zr & \citet{Sing03}, \citet{Bett10}, \citet{Ansa17},\\
              &\citet{Singh18}, \citet{Wern18}, \citet{Witt18},\\
              &\citet{Lher94}, \citet{Urba17}\\
      \hline
      $^{100}$Zr & \citet{Sing08}, \citet{Ansa17}, \citet{Hwan06},  \citet{Urba19a}, \citet{Urba19b}\\
      \hline
      $^{102}$Zr & \citet{Defr09}, \citet{Ansa17}, \citet{Urba19b}\\
      \hline
      $^{104}$Zr & \citet{Blac07}, \citet{Hwan06}, \citet{Brow15}, \\
              & \citet{Hotc90}, \citet{Yeoh10}\\
      \hline
      $^{106}$Zr & \citet{Sing15}, \citet{Sumi11}, \citet{Navin14}, \citet{Brow15}\\
      \hline
      $^{108}$Zr & \citet{Sing15b}, \citet{Sumi11}, \citet{Kame12}\\
      \hline
      $^{110}$Zr & \citet{Paul17}\\
      \hline
    \end{tabular}
%  \end{ruledtabular}
\end{center}
\end{table}

\begin{figure}[hbt]
 \centering
 \includegraphics[width=0.60\linewidth]{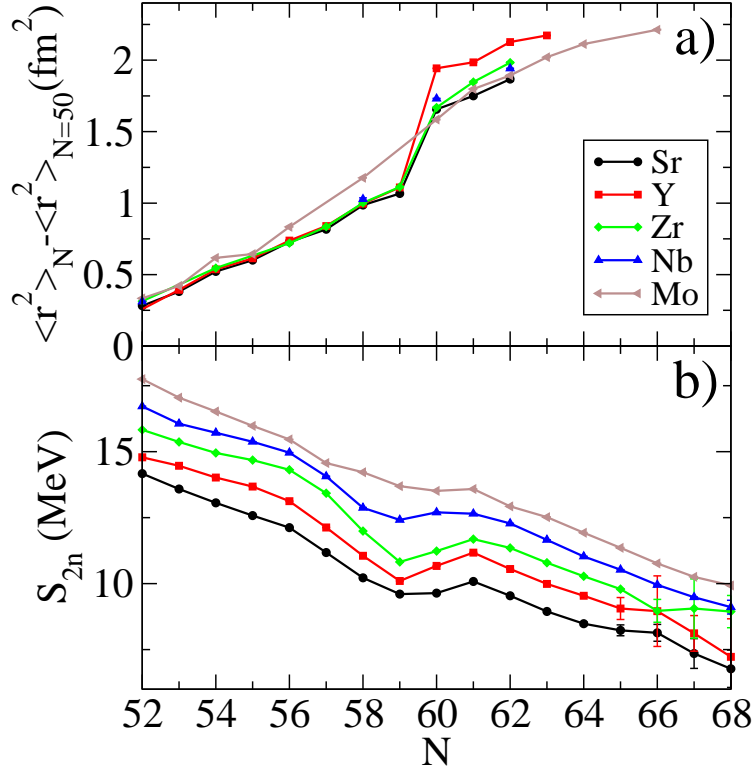}
\caption{Experimental radii (a) and two-neutron separation energies (b) for Sr, Y, Zr, Nb, and Mo.}
\label{fig-s2n-radii}
\end{figure}

Having concentrated on energy spectra, it is of major importance to present at the very start those data that highlight observables indicating both the radii in the Zr nuclei (see, e.g., Fig.\ 26 in \cite{heyde11}, in which also $\beta{_2}$ values are presented), as well as a plot with the S$_{2n}$ values. This information is presented in Fig.\ \ref{fig-s2n-radii} where the radii (panel (a)) and the two-neutron separation energy (panel (b)) are depicted for Sr, Y, Zr, Nb, and Mo isotopes. This figure clearly points towards a sudden change of structure at $N=60$, being specially sharp for Sr, Y, and Zr, smoother in the case of Nb and being almost washed out for Mo.

\section{Theoretical framework}
\label{sec-theo}

\subsection{Shell-model studies: from early works to state-of-the art calculations}
\label{sec-SM}
The early experimental evidence for an unexpected new region of deformation in the A  $\approx$~100 mass region came through the work on primary fission products of $^{252}$Cf, covering both the Sr, Zr, Mo, Ru and Pd nuclei with respect to both energy spectra as well as $B(E2; 2^+_1 \rightarrow 0^+_1)$  reduced transition probabilities \cite{Cheif70}.  Detailed early shell-model calculations were carried out by \citet{Talmi60,Talmi62} using a set of 1d$_{5/2}^n$ configurations with a short-range interaction emphasizing the importance of the proton-neutron T=0 effective interaction \cite{Unna63}. The underlying idea had already been proposed about 10 years before by \citet{Shalit53} with respect to collective effects, e.g., the lowering of the 2$^+_1$ excited state found its origin through the strong mixing of shell-model configurations, and was connected to a very large overlap of the radial proton and neutron wave functions with the constraint of equal radial and orbital quantum numbers.

It was the extensive work of Federman, Pittel and co-workers who carried out the best that could be done at that time, pointing out a microscopic approach within the framework of the shell-model allowing to study the transition when moving  beyond $N=58$, using a model space consisting of the 3s$_{1/2}$, 2d$_{3/2}$ and 1g$_{7/2}$ neutron orbitals and protons restricted to the  2p$_{1/2}$, 1g$_{9/2}$ and 2d$_{5/2}$ orbitals, with most interesting results.  These results are in line with the calculation of the changing single-particle energies through the self-consistent Hartree-Fock (HF) modification, as pointed out by \citet{Good76,Good77}, and \citet{Zeld83}, emphasising the very importance of the residual  proton-neutron interactions resulting in major modifications in the occupation of the 1g$_{9/2}$ proton and 1g$_{7/2}$ neutron orbitals, thereby favoring the formation of a zone of deformed states in the Zr isotopes beyond $N=58$. This approach, as suggested by Federman and Pittel in 1979, entitled as ``a unified shell-model description of nuclear deformation'', emphasizing the importance of the simultaneous occupation of neutrons and protons of spin-orbit partners \cite{Fede77,Fede79a,Fede79b,Fede84,Heyd88,Etch89,Pitt93}, is at the origin of a number of present-day calculations related to the variation of the single-particle energies of nucleons in nuclei, as well as the interplay between the way the nucleon re-distribution of states over the single-particle model space basis and the effective NN force can accommodate in a natural way the appearance of new and unexpected regions of deformation.     

This early combination of various correlation effects and their interplay points towards the final outcome: the importance of the changing mean field as a function of neutron (proton) number for a given isotope (isotone) series of nuclei. One sometimes notices a sensitive final result as a competition between, on one side, the cost to construct a non-standard distribution of nucleons over the available proton (neutron) single-particle orbitals and, on the other hand, the energy gain caused by, in particular, the low-multipole energy correlation gain resulting in a specific energy dependence as a function of neutron number (proton number) outside of closed shells (see \cite{Hey87} where it was shown that in the Zr region this competition was giving rise to important changes in the ground-state $0^+$ structure). 

More recently, detailed LSSM calculations have covered within large model spaces these effects in a more realistic way, in particular, by \citet{Holt00}, covering the $N=50-58$ region, as well as the later shell-model study by \citet{Sieja09}, covering both even-even and odd-mass Zr nuclei in the region from $N=50$  up to $N=58$. The above calculations result in the explanation of the specific energy variation for the first excited $2^+$ level as well as presenting a comparison with the known $E2$ reduced transition probabilities.  Poves put forward an idea  on what might be happening in the Sr, Zr region on what he calls "entanglement" \cite{Poves16}.  He suggested that, resulting from the specific characteristics observed in the mixing of a reference closed-shell nucleus with multi-particle multi-hole excitations across the closed shell, such as 2p-2h, 4p-4h up to 6p-6h excitations as shown in the case of $^{32}$Mg, a particular mixing pattern shows up pointing towards the mixing between shell-model 2p-2h and 4p-4h configurations, giving rise to a mixing of a deformed and a superdeformed shape. Because of the two-body nature of the interaction, no direct coupling between, e.g., a spherical 0p-0h (deformed) and 4p-4h (superdeformed) is allowed. However, detailed shell-model calculations have indicated the possibility of a spherical ground state (0p-0h) configuration mixing mainly with the 4p-4h configuration, even though no direct mixing is possible. This particular structure seems to be observed in the nearby Kr nuclei, with $^{72}$Kr as an example stressing the strange coupling between a spherical and a superdeformed shape, resulting in a final 0$^+$ state called ``a shape entangled'' state. More details are discussed by \citet{Poves16}.  

The need to include a number of adjacent harmonic oscillator shells, compared to the single harmonic oscillator shell studies in order to allow for mp-nh excitations with respect to the more simplified approach quickly runs in too large dimensions, in particular, if one aims to cover the medium-heavy and heavy nuclei. Recent developments going under the name of Monte Carlo shell model, by \citet{Otsuka01,Shimizu12, Shimizu17} (MCSM), have reached the computing power to study nuclei in the lead mass region, which is well known for its  spectacular examples of shape coexistence over a large span of isotopes, in  Pb, Po, Hg, and Pt nuclei.  The Zr region has been another particular region of interest near $N=60$, where the ground-state structure is undergoing a major change (see \cite{Togashi16}) with strong increase, in particular, for the $B(E2; 2^+_1 \rightarrow 0^+_1)$ reduced transition probability. This mass region has been a particularly important one in order to scrutinize the nuclear structure properties using various complementary experimental tools such as lifetime measurements, Coulomb excitation and inelastic electron scattering. Of course, transfer reactions, in particular two-particle transfer, do play a major role in that respect as they are very sensitive to the occupation of the underlying single-particle mean-field orbitals, thereby making detailed tests feasible. In this respect, it is worth to mention the recent IBM works \cite{Vittu18a,Vittu18b}.  Very much the same physics content, albeit within a restricted space, emphasizing the interplay between the effects of a changing cost in energy to redistribute the protons from the 1g$_{9/2}$ filled orbital into orbitals above the $Z=50$ closed shell (unperturbed energy cost to create 2p-2h excitations, reduced through the extra pairing energy, besides including the monopole shift for these 2p-2h excitations, as well as, the major and systematic energy gain through the increased proton-neutron correlations), gave rise \cite{Hey85,Hey87} to an expected  crossing of the regular ground-state for the lower mass values with a redistributed proton occupation at $N \sim 60$.

There is a particular interesting situation appearing in the $^{96}$Zr nucleus with respect to the increase in the excitation energy of the 2$^+_1$ state, as compared with both the lighter and heavier mass Zr isotopes, with a strong drop in energy at $A=98$ and the appearance of a transition into a different nuclear structure regime where deformation has been shown to play a major role. This is a point on which, with the present status in the experimental data, no clear cut answer is available as to the precise microsocopic origin for this behavior. In the paper of \citet{Togashi16}, only proton occupation probabilities are given, but neutron occupation ones are equally important. The already mentioned shell-model calculations by \citet{Holt00} and \citet{Sieja09} with different model spaces, treating the 2p$_{1/2}$, 1g$_{9/2}$ proton shells and the full $50-82$ neutron model space in the former study ($N=52$ up to $N=60$ nuclei) and using the larger proton model space in the latter, in which the $^{58}$Ni nucleus is used as the core nucleus, leading to the use of the proton 1f$_{5/2}$, 2p$_{1/2}$, 2p$_{3/2}$, 1g$_{9/2}$ orbitals, as well as the full $50-82$ model space when treating the $N=50$ up to $N=58$ Zr isotopes, reach very much the same results.

The interesting point is that in both papers, the occupation numbers for the different proton and neutron orbitals as a function of neutron number are presented. The calculations of \citet{Sieja09} obtain clear indications for the fact that for the 0$^+_{1,2}$ states, the occupation of the proton 1g$_{9/2}$ orbital has on average $\sim 1.5$ as occupation, with an increase in $^{96}$Zr up to $\sim 2$, while for the neutron 3s$_{1/2}$ orbital the occupation for the $0_2^+$ state is increasing from $\sim 0.1$ in $A=92-94$ up to $\sim 0.6$ and $0.7$ for $A=96$ and $98$, respectively. This hints to a situation that ressembles what happens in $^{90}$Zr (see Fig.\ 50 in \cite{heyde11}) where the bump up in the 2$^+_1$ excitation energy largely results because of a depression in the energy of the unperturbed lowest 0$^+$ configuration.  This is a typical situation that arises whenever a high-j orbital and a low-spin $j=1/2$ orbital appear next to each other which happens in particular near singly-closed shell nuclei. This implies the presence of 2p-2h (and even more complex repartitions) within the shell-model context, implying the introduction of specific correlations that can be at the basis of making collective effect possible in these Zr nuclei, as was pointed out already implicitly in the early work of \citet{Fede77,Fede79a,Fede79b}. Two-neutron and $\alpha$ particle transfer data for the Zr isotopes, as described in Fig.~29 of Ref.\cite{heyde11}, are indicative of the presence of pairing collectivity for both the $0^+_1$ ground state as well as for the first excited 0$^+_2$ state, consistent with the results from shell-model calculations \cite{Holt00,Sieja09} as indicated by the respective proton and neutron occupation numbers in the various orbitals. One observes an almost equal population into $^{96}$Zr and onwards, even a dominant contribution to the 0$^+_2$ state in $^{98}$Zr.  In this respect, (2n, $\alpha$) transfer reactions are of major importance in elucidating the precise decomposition of the wave functions for the Zr isotopes. 

In view of the increasing collective contribution to the observed low-lying excited state structure moving into the neutron-rich region when approaching $A=98$ and beyond, attempts have been made using the Interacting Boson model, including the possibility of including mp-nh excitations (IBM-CM) to study the Zr. For example in \cite{Garc05} such a study has been carried out using only a single configuration,  concentrating on the low-energy spectra and, more in particular, in the description of the two-neutron separation energy \cite{Naimi10,Rinta04}.

\subsection{Mean-field studies: five-dimensional collective Hamiltonian, general Bohr Hamiltonian and generator coordinate method using various forces} 
\label{sec-MF}
The mass $A\approx 100$ region has been extensively studied using self-consistent methods, in particular, concentrating on the way changes in the nuclear overall shape properties have been observed in the Kr, Sr, Zr, and Mo isotopes.

There are various approaches, such as the early HFB method, which resulted in what was called the complex Excited Vampir variational approach, and effective interaction starting from the Bonn A potential, within a large model space \cite{Petr11}, covering in particular the $A=98-110$ Zr isotopes, mainly concentrating on the energy spectra and its deformation characteristics.  

Extensive discussions concerning shape evolution in the Kr, Zr and Sr isotopes have been carried out using covariant density functional theory (CDFT) within the relativistic HFB picture. \citet {Nomu16} also showed that starting from the Gogny energy density functional within the framework of the HFB theory, an extensive mapping of the total energy surfaces for the Zr isotopes, spanning the mass interval $N=54$ ($A=94$) up to $N=70$ ($A=110$), indicated an onset of deformation, starting at $A=98$, moving through a triaxial region with a deformed minimum developing from $A=102$ and onwards up to mass $A=110$. No spectroscopic data are calculated but the mean-field is used to construct an IBM-CM Hamiltonian through a mapping of the energy surface onto the IBM-CM energy surface.

Calculations using a generalized Bohr Hamiltonian, constructed through a reduction of the Hill-Wheeler generator coordinate method (GCM) equation using a Gaussian Overlap Approximation (GOA) by \citet{Dela10} have been carried out for a larger part of the nuclear mass table, including the Zr isotopes. Calculations starting from HFB angular momentum projected states, solving the Hill-Wheeler equations, allow to calculate the  energy surfaces and energy spectra can be calculated over the full ($\beta,\gamma$) plane and have been tested in particular in the Mg region as proof of principal. These exact studies have been used and the results for the Zr nuclei have been presented for the Zr isotopes, as in the paper by \citet{Paul17}, comparing the experimental data with the three different theoretical approaches and coming to the conclusion that the GOA reduction to a five-dimensional Bohr Hamiltonian (5DCH), with the solutions of solving the Hill-Wheeler equations, or using somewhat different approaches in solving the Hill-Wheeler equations, by  \citet{Bender08} using the SLyMR00 version of the Skyrme force \cite{Bender03,Bender08, Bally14, Sadou13}), as well as by  \citet{Borr15,Rod10}, using the Gogny D1S force, exhibit an overall similar variation on the 2$^+_1$ and the R$_{42}$ ratio (E($4_1^+$)/E($2^+_1$)) for the whole Zr isotopic series known at present. Still, some differences show up, in particular for the R$_{42}$ ratio. 

Recently, constrained Hartree-Fock calculations have been carried out for the whole sequence of the Zr isotopes, spanning from $N=40$ up to the $N=82$ neutron closed shell in which indications for a sequence is obtained, first of spherical shape up to $N=56$, with the onset of prolate deformation moving up to $N=72$, changing again into oblate shapes at $N=74$, coming back to spherical shapes approaching the $N=82$ closed shell \cite{Matsu16}. Here, attention has been given to the specific effect that the tensor component of the interaction used in the present study has on the shape evolution. 
Moreover, a recent HF calcution for $^{86-122}$Zr can be found in \cite{Miya18}, where they conclude that nuclei with a spherical shape are obtained for $^{86-96}$Zr, a prolate deformation is obtained for  $^{98-112}$Zr, an oblate shape appears in $^{114-120}$Zr and, finally, for  $^{122}$Zr the spherical shape is recovered. 
   
In view of more recent increased computing power, full symmetry conserving configuration mixing calculations of deformed intrinsic states have been carried out by \citet{Rodr10} within the HFB framework, allowing to handle both the axial and triaxial degrees of freedom on equal footing and with application to the heavier Zr isotopes ($A\approx 110$) \cite{Borr15}. We should mention the efforts carried out in order to obtain a microscopic derivation of a quadrupole Hamiltonian, with particular attention given to the issue of shape coexistence as well as the mixing dynamics by \citet{Matsu16}, with applications to the Kr, Sr and Zr region. 
%TRIAXXXXXXXXXXXXXXXXXXXXXX

It is clear that very large efforts have been invested to explore the power of self-consistent beyond mean-field calculations, allowing to uncover some of the more salient features on nuclear shapes, indicating a transition into a region of strongly deformed nuclei (see Section \ref{sec-exp} on data and Fig.~\ref{fig-e-systematics-zr} with the energy spectra systematics). 
  
\section{The Interacting Boson Model with configuration mixing formalism}
\label{sec-ibm-cm}
\subsection{The formalism}
\label{sec-formalism}

The IBM-CM is a natural extension of the original IBM proposed by \citet{iach87} that is ideal to treat simultaneously and in a consistent way  several boson configurations which correspond to different particle-hole (p-h) excitations across the shell closure \cite{duval81,duval82}. In the particular case of Zr isotopes, the model space includes the regular proton 0p configurations, assuming $Z=40$ as a closed shell, and a number of valence neutrons outside of the $N=50$ closed shells (this is the standard IBM treatment for the Zr even-even nuclei) as well as the proton 2h-2p configurations  and the same number of valence neutrons corresponding to an extended $[N]\oplus[N+2]$ boson space, where $N$ is the boson number, corresponding to the number of active protons outside the $Z=40$, zero in the case of Zr, plus the number of valence neutrons outside the $N=50$ closed shell, counting both proton holes and particles, divided by $2$. Since such fermion excitations across the $Z=40$ proton sub-shell closure, considered in the present IBM approach are treated as genuine s and d bosons, this implicitly assumes that the lowest S ($0^+_1$) two-hole shell-model configuration as well as the lowest D ($2^+_1$) shell-model configuration have the microscopic structure of the used s and d proton-like bosons. Therefore, the complete Hamiltonian, assuming two different sectors in the boson space can be written as,
\begin{equation}
  \hat{H}=\hat{P}^{\dag}_{N}\hat{H}^N_{\rm ecqf}\hat{P}_{N}+
  \hat{P}^{\dag}_{N+2}\left(\hat{H}^{N+2}_{\rm ecqf}+
    \Delta^{N+2}\right)\hat{P}_{N+2}\
  +\hat{V}_{\rm mix}^{N,N+2}~,
\label{eq:ibmhamiltonian}
\end{equation}
where $\hat{P}_{N}$ and $\hat{P}_{N+2}$ are projection operators onto the $[N]$ and the $[N+2]$ boson spaces, respectively, $\hat{V}_{\rm mix}^{N,N+2}$  describes the mixing between the $[N]$ and the $[N+2]$ boson subspaces, and
\begin{equation}
  \hat{H}^i_{\rm ecqf}=\varepsilon_i \hat{n}_d+\kappa'_i
  \hat{L}\cdot\hat{L}+
  \kappa_i
  \hat{Q}(\chi_i)\cdot\hat{Q}(\chi_i),
  \label{eq:cqfhamiltonian}
\end{equation}
is a simplified form of the general IBM Hamiltonian called extended consistent-Q Hamiltonian (ECQF) \cite{warner83,lipas85} with $i=N,N+2$, being $\hat{n}_d$ the $d$ boson number operator, 
\begin{equation}
  \hat{L}_\mu=[d^\dag\times\tilde{d}]^{(1)}_\mu ,
\label{eq:loperator}
\end{equation}
the angular momentum operator, and
\begin{equation}
  \hat{Q}_\mu(\chi_i)=[s^\dag\times\tilde{d}+ d^\dag\times
  s]^{(2)}_\mu+\chi_i[d^\dag\times\tilde{d}]^{(2)}_\mu~,
\label{eq:quadrupoleop}
\end{equation}
the quadrupole operator. This approach has been proven to be a good approximation in several recent papers on Pt \cite{Garc09,Garc11,Garc12}, Hg \cite{Garc14b,Garc15b} and Po isotopes \cite{Garc15,Garc15c}. 

The parameter $\Delta^{N+2}$ can be understood as the energy needed to excite two proton particles across the $Z=40$ shell gap, giving rise to 2p-2h excitations, corrected with the pairing interaction gain and including monopole effects \cite{Hey85,Hey87}. The operator $\hat{V}_{\rm mix}^{N,N+2}$ describes the mixing between the $N$ and the $N+2$ configurations and is defined as
\begin{equation}
  \hat{V}_{\rm mix}^{N,N+2}=\omega_0^{N,N+2}(s^\dag\times s^\dag + s\times
  s)+\omega_2^{N,N+2} (d^\dag\times d^\dag+\tilde{d}\times \tilde{d})^{(0)}.
\label{eq:vmix}
\end{equation}
Note that it is customary to consider the two coefficients as equal,  $\omega_0^{N,N+2}=\omega_2^{N,N+2}=\omega$.

The $E2$ transition operator for two-configuration mixing is defined as the sum of the contribution for both sectors,
\begin{equation}
  \hat{T}(E2)_\mu=\sum_{i=N,N+2} e_i \hat{P}_i^\dag\hat{Q}_\mu(\chi_i)\hat{P}_i~,
  \label{eq:e2operator}
\end{equation}
where the $e_i$ ($i=N,N+2$) are the effective boson charges and $\hat{Q}_\mu(\chi_i)$ the quadrupole operator defined in equation (\ref{eq:quadrupoleop}), therefore being the same in the Hamiltonian (\ref{eq:cqfhamiltonian}) and in the $\hat{T}(E2)$ operator (\ref{eq:e2operator}). Note that every term only affects to a given sector and there are no crossing contributions between both sectors.

The former operators, Hamiltonian and $\hat{T}(E2)$ operator, present a set of free parameters that should be fixed following the procedure that will be described in section \ref{sec-fit-procedure}.

Once the Hamiltonian is diagonalized, the resulting wave functions, within the IBM-CM, can be written as 
\begin{eqnarray}
\Psi(k,JM) &=& \sum_{i} a^{k}_i(J;N) \psi((sd)^{N}_{i};JM) 
\nonumber\\
&+& \
\sum_{j} b^{k}_j(J;N+2)\psi((sd)^{N+2}_{j};JM)~,
\label{eq:wf:U5}
\end{eqnarray}
where $k$ refers to the different states with a given $J$, while $i$, and $j$ run over the bases of the $[N]$ and $[N+2]$ sectors, respectively. The  weight of the wave function contained within the $[N]$-boson subspace, can then be defined as the sum of the squared amplitudes,
\begin{equation} 
  w^k(J,N) \equiv \sum_{i}\mid a^{k}_i(J;N)\mid ^2.
  \label{wk}
\end{equation} 
Likewise, one obtains the content in the $[N+2]$-boson subspace.

\subsection{The fitting procedure: energy spectra and absolute $B(E2)$ reduced transition probabilities}
\label{sec-fit-procedure}

In this Section we explain in detail how the parameters of the Hamiltonian (\ref{eq:ibmhamiltonian}), (\ref{eq:cqfhamiltonian}), (\ref{eq:quadrupoleop}),  and (\ref{eq:vmix}) and the effective charges in the $\hat{T}(E2)$ transition operator (\ref{eq:e2operator}) have been determined.

We study the range $^{94}$Zr to $^{110}$Zr, thereby, covering almost the whole first half of the neutron shell $N=50-82$ and part of the second. Note that we do not consider the isotopes $^{90-92}$Zr because they are too close to the shell closure, $N=50$, and, therefore, the IBM calculations are not reliable enough.
\begin{table}
\caption{Energy levels, characterized by $J^{\pi}_i$, included in the energy fit for the case of $^{94}$Zr and the assigned $\sigma$ values in keV. A similar set of levels, but not the same, is adopted for each isotope.}
\label{tab-energ-fit}
\begin{center}
  \begin{tabular}{|c|c|}
    \hline
    Error (keV) & States  \\
    \hline
    $\sigma=1$& $2_1^+$ \\
    $\sigma=10$ & $4_1^+, 0_2^+, 2_2^+, 4_2^+$\\
    $\sigma=100$  & $2_3^+, 2_4^+, 3_1^+, 4_3^+, 4_4^+$\\
    \hline
  \end{tabular}
\end{center}
\end{table}

In the fitting procedure carried out here, we try to obtain the best overall agreement with the experimental data including both the excitation energies and the $B(E2)$ reduced transition probabilities. Using the expression of the IBM-CM Hamiltonian, as given in equation (\ref{eq:ibmhamiltonian}), and of the $E2$ operator, as given in equation (\ref{eq:e2operator}), in the most general case $12$ parameters are necessary.  We impose as a constraint to obtain parameters  that  change smoothly in passing from isotope to isotope. Moreover, we try to keep as many parameters as possible at a constant value. However, the closeness of the $N=50$ shell closure and the subshell closure  for $N=56$, related with the filling of the 2d$_{5/2}$ orbital, strongly perturbs the value of several parameters. As a matter of fact, the value of the mixing part, $\omega$, cannot be kept constant, but it changes for $A=94$ ($\omega=150$ keV) with respect to the rest of the isotope chain ($\omega=15$ keV), and the value of  $\Delta^{N+2}$ also evolves, being larger for the lightest isotopes, around $\Delta^{N+2}=3200$ keV and it drops down to $\Delta^{N+2}=820$ keV for the majority of isotopes. A microscopic study of the variation in the energy cost to move nucleons from normal ordering into a 2p-2h excitation across the $Z=40$ closed shell has been performed basing on the combined effect of the energy gap, the pairing energy and the monopole shift in the proton single-particle energies as shown in Figs.\ 14 and 15 of \cite{Hey87}. It explains the drop in the value of $\Delta$ as used in the present study.  Besides, the coefficients $\kappa'_N$ and $\kappa'_{N+2}$ can be taken to vanish for $^{100-110}$Zr, but they result in a non-zero value for $^{94-98}$Zr. 
\begin{table}
\caption{Hamiltonian and $\hat{T}(E2)$ parameters resulting from the present study.  All quantities have the dimension of energy (given in keV), except $\chi_{N}$ and $\chi_{N+2}$, which are dimensionless and $e_{N}$ and $e_{N+2}$ which are given in units $\sqrt{\mbox{W.u.}}$. The used number of bosons for the regular and the intruder sector, respectively, is indicated in the first column.}
\label{tab-fit-par-mix}
\begin{center}
\begin{ruledtabular}
\begin{tabular}{ccccccccccccc}
Nucleus (N,N+2)&$\varepsilon_N$&$\kappa_N$&$\chi_{N}$&$\kappa'_N$&$\varepsilon_{N+2}$& $\kappa_{N+2}$&$\chi_{N+2}$&$\kappa'_{N+2}$& $\omega$& $\Delta$&$e_{N}$&$e_{N+2}$\\
\hline
 $^{94}$Zr (2,4)&   1201&    -0.00&  1.30&  -39.93&     0.1&   -26.32& -2.35&   21.97& 150&   3200 &  2.01&  -1.36 \\
 $^{96}$Zr (3,5)&   1800&   -34.41&  1.82&   25.12&   333.2&   -29.18&  0.09&   -4.50&  15&   2000 &  0.90&   3.35 \\
 $^{98}$Zr (4,6)&   1044&   -25.23&  1.80&   78.71&   439.6&   -14.32&  0.67&   26.48&  15&    814 &  1.55&   3.11 \\
$^{100}$Zr (5,7)&   1063&   -23.26&  2.53&    0.00&   438.3&   -28.76& -0.95&    0.00&  15&    820 &  0.46&   2.26 \\
$^{102}$Zr (6,8)&   1050&   -23.58&  2.46&    0.00&   337.9&   -32.01& -0.68&    0.00&  15&    820 &  0.46&   2.32 \\
$^{104}$Zr (7,9)&   1050\footnotemark[1]&   -23.58\footnotemark[1]&  2.46\footnotemark[1]&    0.00\footnotemark[1]&   616.5&   -32.00& -1.35&    0.00&  15&    820 &  0.46\footnotemark[1]&   2.04 \\
$^{106}$Zr (8,10)&   1050\footnotemark[1]&   -23.58\footnotemark[1]&  2.46\footnotemark[1]&    0.00\footnotemark[1]&   580.5&   -31.03& -0.93&    0.00&  15&    820 &  0.46\footnotemark[1]&   1.79 \\
$^{108}$Zr (7,9)&   1050\footnotemark[1]&   -23.58\footnotemark[1]&  2.46\footnotemark[1]&    0.00\footnotemark[1]&   540.2&   -30.00& -0.90&    0.00&  15&    820 &  0.46\footnotemark[1]&   1.79\footnotemark[2] \\
$^{110}$Zr (6,8)&   1050\footnotemark[1]&   -23.58\footnotemark[1]&  2.46\footnotemark[1]&    0.00\footnotemark[1]&   498.9&   -32.00& -0.90&    0.00&  15&    820 &  0.46\footnotemark[1]&   1.79\footnotemark[2] \\
\end{tabular}
\end{ruledtabular}
\end{center}
\footnotetext[1]{Hamiltonian and $\hat{T}(E2)$ parameters for the regular sector taken from $^{102}$Zr.}
\footnotetext[2]{Effective charge for the intruder sector taken from $^{106}$Zr.}
\end{table}

The $\chi^2$ test is used in the fitting procedure in order to extract the optimal solution. The $\chi^2$ function is defined in the standard way as
\begin{equation}
  \label{chi2}
  \chi^2=\frac{1}{N_{data}-N_{par}}\sum_{i=1}^{N_{data}}\frac{(X_i(data)-X_i (IBM))^2}{\sigma_i^2},
\end{equation} 
where $N_{data}$ is the number of experimental data, $N_{par}$ is the number of parameters used in the IBM fit, $X_i(data)$ describes the experimental excitation energy of a given excited state (or an experimental $B(E2)$ value), $X_i(IBM)$ denotes the corresponding calculated IBM-CM value, and $\sigma_i$ is an error (theoretical) assigned to each $X_i(data)$ point. We minimize the $\chi^2$ function for each isotope separately using the package MINUIT \cite{minuit} which allows to minimize any multi-variable function.

As input values,  we start from the excitation energies of the levels corresponding to those levels presented in Table \ref{tab-energ-fit}, although only the particular case of $^{94}$Zr is presented. In this table we also give the corresponding $\sigma$ values. In the heavier Zr isotopes, a similar set of states is considered, although in every isotope differences are established, concerning the considered states and the value of $\sigma$, in order to allow a smooth convergence of the fitting procedure. We stress that the $\sigma$ values do not correspond to experimental error bars, but they are related with the expected accuracy of the IBM-CM calculation to reproduce a particular experimental data point. Thus, they lead the calculation into a smooth convergence towards the corresponding experimental level. Note that we have only considered states in the fit with angular momentum and parity that are unambiguously determined, unless they are assumed to {\it do} not exhibit a collective behavior.

For the $E2$ transitions rates, we have used the available experimental data involving the states presented in Table \ref{tab-energ-fit}, but not only, restricted to those $E2$ transitions for which absolute $B(E2)$ values are known, except if serious hints that the states involved present non-collective degrees of freedom exist. Additionally, we have taken a value of $\sigma$ that corresponds to $10\%$ of the $B(E2)$ values or to the experimental error bar if larger, except for the transition $2_1^+\rightarrow 0_1^+$ where a smaller value of $\sigma$ ($0.1$ W.u.) was taken, thereby normalizing, in most of cases, our calculated values to the experimental $B(E2;2^+_1 \rightarrow 0^+_1)$ value. All the considered transitions can be found in Table \ref{tab-be2a}. 

The resulting values of the parameters for the IBM-CM Hamiltonian and $\hat{T}(E2)$ operator, are given in Table \ref{tab-fit-par-mix}. In this Table certain parameters could not be determined from the experimental information. In particular, the parameters corresponding to the regular sector of $^{104-110}$Zr because all known experimental states belong to the intruder sector, therefore, we have assumed to use the same values as for $^{102}$Zr. The same happens for the intruder effective charge of $^{108-110}$Zr which was taken from $^{106}$Zr.  
\begin{figure}
  \centering
  \includegraphics[width=0.5\linewidth]{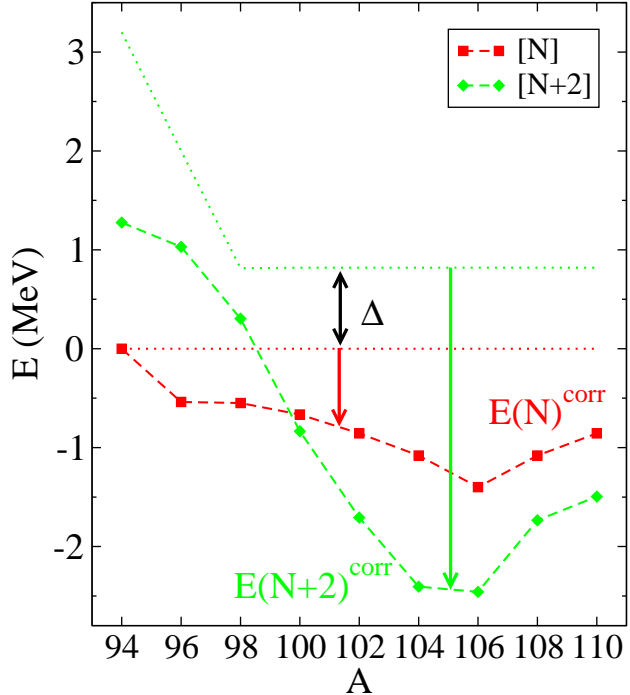}
  \caption{Absolute energy of the lowest unperturbed regular and intruder 0$^+_1$ states for $^{94-110}$Zr. The arrows correspond to the correlation energies in the N and N+2 subspaces (see also the text for a more detailed discussion).}
  \label{fig-energ-corr}
\end{figure}

\subsection{Correlation energy in the configuration mixing approach} 
\label{sec-corr_energy}
The intruder states are expected to appear at energies well above the regular ones with identical angular momentum. The reason is that it is needed to provide the necessary energy to promote a pair of protons across the $Z=40$ shell gap, corrected because of the strong pairing energy gain when forming two extra $0^+$ coupled (particle and hole) pairs. In the case of Zr this value is $\Delta^{N+2}=3200$ keV for $^{94}$Zr but steadily decreases down to $\Delta^{N+2}=820$ keV for $^{100}$Zr and onwards. However, this energy should be strongly  corrected by the quadrupole energy gain when opening up the proton shell, as well as by the monopole correction caused by a change in the single-particle energy gap at $Z=40$ as a function of the neutron number \cite{Hey87}. In particular, around midshell, $N=66$, where the number of active valence nucleons is maximal, one expects to observe a larger correction. In certain isotope chains, the correction is so large that the intruder configuration can become the ground state of the system, as it happens in Pt and Po nuclei around the midshell, while in Hg and Pb, the intruder states never go below the regular configurations.

According to the energy and the transition rate systematics, it is not always simple, or even possible, to assign a well defined character, either regular or intruder, to every state. This is particularly difficult in the case of a strong mixing between both families of states. A possible solution is to use the IBM-CM calculation, once the free parameters have been fixed,  switching off the mixing term in the Hamiltonian, thereby, setting $\omega=0$. Doing so we can calculate explicitly the ``absolute'' energy of the lowest $0^+$ state belonging to both the regular [N] and [N+2] intruder configuration spaces. We use as a reference the energy of a purely vibrational regular Hamiltonian (all the parameters vanish except $\varepsilon_{N}>0$), which corresponds to the horizontal red dotted line in Fig.~\ref{fig-energ-corr}. The energy needed to create the extra pair of bosons across the shell gap is the upper green dotted line in Fig.~\ref{fig-energ-corr}. The energy difference between both lines corresponds to the value of $\Delta^{N+2}$. The energy of the  0$^+$ state in the regular sector, [N], is below the red dotted line, which is a consequence of the correlation energy that lowers the energy with respect to the vibrational situation. The same happens for the case of the lowest 0$^+$ intruder state with respect to the green dotted line. The regular configuration corresponds to a slightly deformed shape, while the intruder one to a much more deformed one. This explains why the correlation energy gained by the intruder state is much larger than the one gained by the regular one. The reason is two 
fold. On one hand, the part of the Hamiltonian acting on the intruder sector presents a much stronger quadrupole interaction as compared with the regular one. Consequently, it induces a larger deformation. On the other hand, the effective number of bosons in the intruder sector is two units larger than that in the regular one, which, as a matter of fact, also contributes to a larger deformation and larger correlation energy in the intruder sector. The relative position of these lowest regular and intruder states is plotted in Fig.~\ref{fig-energ-corr}. This explains why the energies of both configurations cross at $N=100$ and the intruder configuration becomes the lower configuration for this isotope and onwards. Consequently, the relative position of both configurations depends on the balance between the off-set, $\Delta^{N+2}$, and the difference in the correlation energy $E(N+2)^{corr} - E(N)^{corr}$.    

In a detailed way, one can see how the position of the intruder configuration drops rapidly reaching its minimum at the midshell because, on one hand, the off-set, $\Delta^{N+2}$, reduces and, on the other hand, because the number of effective bosons increases. The position of the regular configuration remains rather stable because the Hamiltonian corresponding to this sector generates spherical shapes. In the case of $A=94$, the correlation energy of the intruder configuration is quite large while almost zero for the regular one. For $A=96$ and $98$ both contributions are almost equal. However, from $A=100$ and onwards the correlation energy for the intruder configuration is much larger as compared with the correlation energy for the regular configuration.

\begin{figure}[hbt]
  \centering
  \includegraphics[width=.5\linewidth]{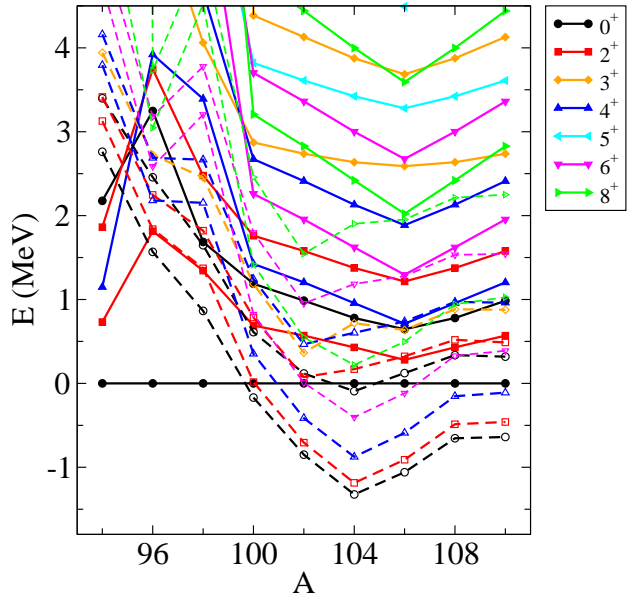}
  \caption{Energy spectra for the IBM-CM Hamiltonian presented in Table \ref{tab-fit-par-mix}, switching off the mixing term. The two lowest-lying regular states and the lowest-lying intruder state for each of the angular momenta are shown (full lines with closed symbols for the regular states while dashed lines with open symbols are used for the intruder ones).}
  \label{fig-ener-nomix}
\end{figure}

To complement the above analysis, we can also consider the energy systematics for the complete spectra.
The uncoupled N and N+2 boson systems, as shown in Fig.~\ref{fig-ener-nomix}, are extremely valuable to highlight whether a parabolic behavior is obtained for the intruder state energy systematics. In Fig.~\ref{fig-ener-nomix} the energy systematics for levels below $4.5$ MeV is depicted, considering two regular and two intruder states per angular momentum. As it is customary, the energies are referred to the energy of the first $0^+$ regular state. In this figure, the regular states (full lines) present a large kink at $A=96$ due to the $d_{5/2}$ subshell filling, already mentioned  before, and then a much more flat behavior from $A=100$ and onwards, although showing a small dip right at midshell, $A=106$. Concerning the intruder states (dashed lines), they exhibit the expected parabolic behavior, although, in this case, the minimum of the energy happens at $A=104$ and not at midshell and, moreover, instead of a smooth parabola it presents an inverted peak. The reason is once more to be seen in Fig.~\ref{fig-energ-corr}, where the regular states indicate a rapid decrease moving from $A=104$ to $106$ as compared with the intruder states. Note that most probably this fact has been generated because of having taken the same Hamiltonian parameters in the regular sector for $A=104-110$ than for $A=102$, because the lack of experimental information. 

\vspace*{-1cm}
\begingroup
\squeezetable
\begin{table}
  \caption{Comparison of the experimental absolute $B(E2)$ values (given in W.u.) with the IBM-CM Hamiltonian results. Data are taken from the corresponding  Nuclear Data Sheets \cite{Abri06,Abri08,Sing03,Sing08,Defr09,Blac07,Sing15,Sing15b}, complemented with references given in the table footnotes.}  
  \label{tab-be2a}
  \vspace{-.5cm}
  \begin{center}
\begin{ruledtabular}
\begin{tabular}{cccc}
Isotope   &Transition             &Experiment&IBM-CM \\
\hline
  $^{94}$Zr&$2_1^+\rightarrow 0_1^+$& 4.9(3)         & 4.9\\  
          &$0_2^+\rightarrow 2_1^+$& 9.4(4)       & 10.6 \\   
          &$4_1^+\rightarrow 2_1^+$& 0.879(23)       & 1.7 \\   
          &$2_2^+\rightarrow 0_2^+$& 19(2)\footnotemark[1]        & 21 \\   
          &$2_2^+\rightarrow 2_1^+$& 0.06($^{+0.13}_{-0.06}$)\footnotemark[1]        & 0.05 \\   
          &$2_2^+\rightarrow 0_1^+$& 3.9(3)\footnotemark[1]        & 1.9 \\   
          &$4_2^+\rightarrow 2_2^+$& 34($^{+10}_{-17}$)\footnotemark[1]       & 13 \\   
          &$4_2^+\rightarrow 2_1^+$& 13($^{+4}_{-7}$)\footnotemark[1]       & 9 \\   
\hline
  $^{96}$Zr&$2_1^+\rightarrow 0_1^+$& 2.3(3)         &2.3    \\  
          &$3_1^+\rightarrow 2_1^+$& 0.1($^{+0.3}_{-0.1}$)       & 0.06 \\   
          &$2_3^+\rightarrow 2_1^+$& 50(7)       & 46 \\   
          &$0_3^+\rightarrow 2_2^+$& 34(9)       & 16 \\   
          &$4_2^+\rightarrow 2_2^+$& 56(44)       & 0 \\   
          &$4_2^+\rightarrow 2_1^+$& 16($^{+5}_{-13}$)  & 0 \\   
          &$0_4^+\rightarrow 2_1^+$& 0.3(3)       & 0 \\   
          &$0_4^+\rightarrow 2_2^+$& 1.8(14)       & 0 \\   
          &$2_2^+\rightarrow 0_2^+$& 36(11)\footnotemark[2]       & 54 \\   
          &$2_2^+\rightarrow 0_1^+$& 0.26(8)\footnotemark[2]       & 0.5 \\   
\hline
  $^{98}$Zr&$2_1^+\rightarrow 0_1^+$& 2.9(6)\footnotemark[3]  and $\geq 1.83$\footnotemark[4]$^,$\footnotemark[5]        & 9.6    \\  
          &$2_1^+\rightarrow 0_2^+$& 28.3(65)\footnotemark[3]        & 32 \\   
          &$4_1^+\rightarrow 2_1^+$& 43.3(138)\footnotemark[3] and $\geq 20.75$\footnotemark[4]$^,$\footnotemark[5]       & 59 \\   
          &$4_1^+\rightarrow 2_2^+$& 67.5(216)\footnotemark[3]        & 67 \\   
          &$6_1^+\rightarrow 4_1^+$& 103.0(357)\footnotemark[3]  & 143 \\   
          &$0_3^+\rightarrow 2_1^+$& 51(5)  &  53  \\   
          &$0_4^+\rightarrow 2_2^+$& 44(4)  &  42  \\   
          &$0_4^+\rightarrow 2_1^+$& 0.107(14)  &  0.33  \\   
  \hline
  $^{100}$Zr&$2_1^+\rightarrow 0_1^+$& 74(4)         & 70    \\  
          &$0_2^+\rightarrow 2_1^+$& 67(7)       & 64 \\   
          &$4_1^+\rightarrow 2_1^+$& 103(9)       & 120 \\   
          &$6_1^+\rightarrow 4_1^+$& 140(30) & 128 \\   
          &$8_1^+\rightarrow 6_1^+$& 124(13)  &  122  \\   
          &$10_1^+\rightarrow 8_1^+$& 124(15)  &  105  \\   
          &$6_3^+\rightarrow 4_2^+$& 0.011(3)      & 0.05   \\   
          &$6_3^+\rightarrow 4_1^+$& 0.00019(6)     & 0   \\   
  \hline
  $^{102}$Zr&$2_1^+\rightarrow 0_1^+$& 105 (14)         & 109    \\  
          &$4_1^+\rightarrow 2_1^+$& 167$(^{+30}_{-22})$\footnotemark[4]          &  155\\   
  \hline
  $^{104}$Zr&$2_1^+\rightarrow 0_1^+$& 139($^{+11}_{-13}$)\footnotemark[6] and 180(30)\footnotemark[5]     & 139    \\  
  \hline
  $^{106}$Zr&$2_1^+\rightarrow 0_1^+$& 104(3)\footnotemark[6]         & 104     
\end{tabular}
\end{ruledtabular}
\end{center}
\vspace*{-.7cm}
\footnotetext[1]{Data taken from Ref.~\cite{Chak13}.}
\footnotetext[2]{Data taken from Ref.~\cite{Krem16}.}
\footnotetext[3]{Data taken from Ref.~\cite{Singh18}.}
\footnotetext[4]{Data taken from Ref.~\cite{Ansa17}.}
\footnotetext[5]{Experimental data not included in the fit.}
\footnotetext[6]{Data taken from Ref.~\cite{Brow15}.}
\end{table}
\endgroup

\subsection{Detailed comparison between the experimental data and the IBM-CM results:  energy spectra and $E2$ transition rates}
\label{sec-fit-compare}
\begin{figure}[hbt]
  \centering
  \includegraphics[width=0.8\linewidth]{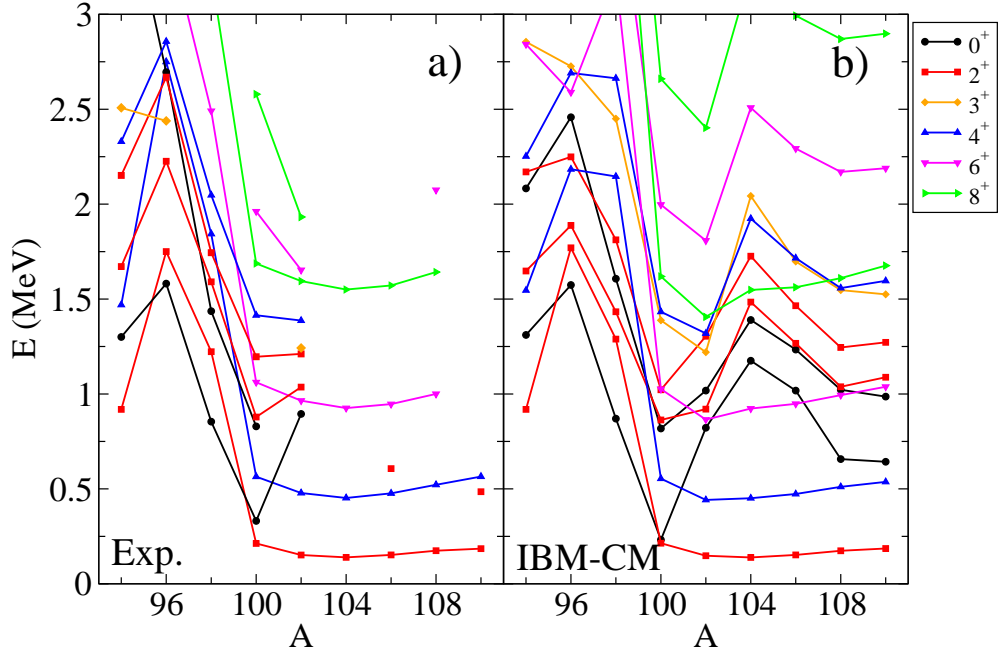}
  \caption{Experimental excitation energies (up to $E \approx 3.0$  MeV) (panel (a)) and the theoretical results (panel (b)),  obtained from the IBM-CM.} 
  \label{fig-energ-comp}
\end{figure}

In this section, we compare in detail the experimental energy spectra with the theoretical ones up to an excitation energy below $E \approx 3.0$ MeV. In Fig.~\ref{fig-energ-comp}a we plot the experimental data set, while in  Fig.~\ref{fig-energ-comp}b the theoretical values are depicted. First, one notices how the theoretical energy of the $2_1^+$ state closely follows the experimental one which is the result of the use of these energies as the normalization energy, resulting from imposing a small value of $\sigma$ for this  particular energy level in the fitting procedure. The energies from $A=94$ to $A=98$ are strongly affected by the subshell closure at $A=96$ which is the reason for the peak appearing at this mass number, and is perfectly reproduced by the theoretical calculations. The drop of the energies down to $A=100$ is due to the onset of deformation which induces a much more compressed spectrum. Once the region $A=104-110$ is reached, only few experimental levels are known, corresponding to intruder states. On the theoretical side, the flat energy systematics is very well reproduced but other states, which are indeed regular ones, do also become visible. The precise position of these regular states should be considered as only qualitative because the regular parts of the Hamiltonian for these isotopes have been considered as identical to $A=102$, because of the lack of experimental information. The IBM-CM results, in this respect, can be used as predictions for the presence of such states. The position of the peak at $A=104$ can be understood by considering the behavior of Fig.~\ref{fig-energ-corr}, where one notices that both curves -the regular and the intruder band head energies- exhibit a minimum at $A=106$, but in the case of the intruder state, the energy for $A=104$ and $106$ is almost the same, while for the regular one, the energy for $A=104$ is well above that of $A=106$. We further emphasize the presence of a second $2^+$ state for $A=106$ and $110$ on the experimental side, states  that do not appear in the theoretical energy spectra. Most probably, these states correspond to 4p-4h excitations that are not considered in the IBM-CM Hilbert space used in  the present paper.

\begin{figure}[hbt]
  \centering
  \includegraphics[width=0.6\linewidth]{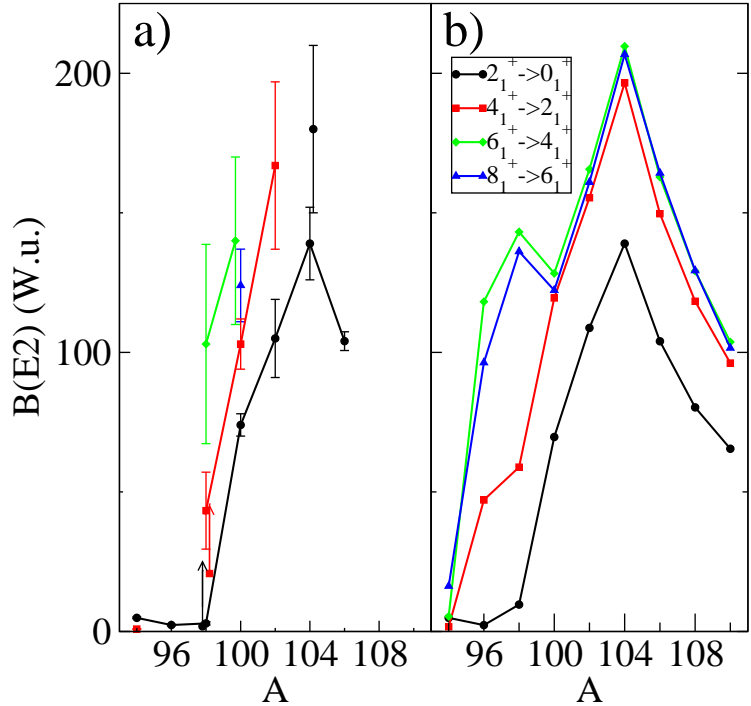}
  \caption{Comparison of the absolute $B(E2)$ reduced transition probabilities along the yrast band, given in W.u. Panel (a) corresponds to known experimental data and panel (b) to the theoretical IBM-CM results.} 
  \label{fig-be2-1}
\end{figure}

\begin{figure}[hbt]
  \centering
  \includegraphics[width=0.6\linewidth]{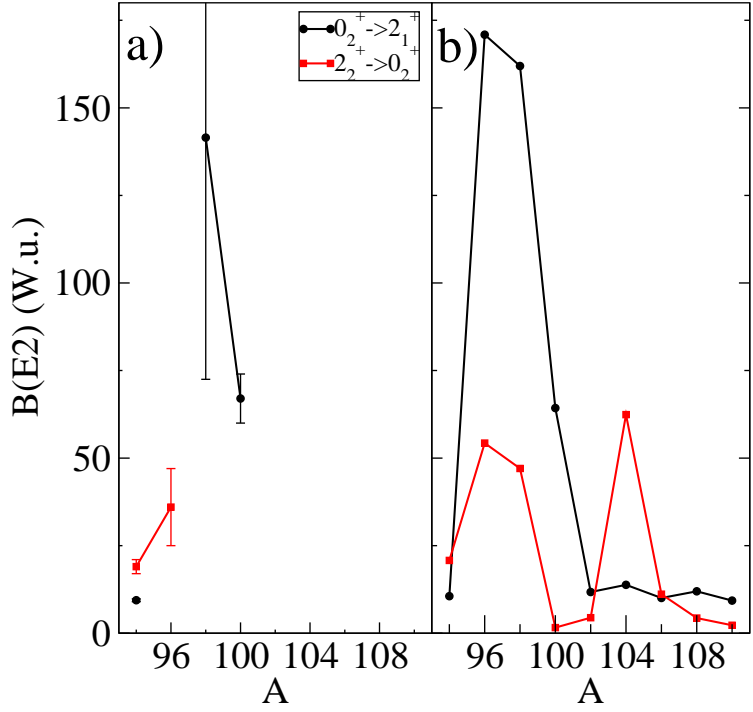}
  \caption{Comparison of few non-yrast intraband  absolute $B(E2)$ reduced transition  probabilities, given in W.u. Panel (a) corresponds to the few known experimental data, panel (b) to the theoretical IBM-CM results.}
  \label{fig-be2-2}
\end{figure}

Next, we carry out a comparison for the $B(E2)$ values, which reveal whether a consistent set of wave functions is generated in the IBM-CM calculation or not. The experimental information is extensive for some nuclei, while scarce for others. However, an extensive experimental program is providing new experimental information (see for example \cite{Singh18}). In Figs.~\ref{fig-be2-1} and \ref{fig-be2-2} we compare the $B(E2)$ reduced transition probabilities. We also present a more detailed comparison on $B(E2)$ values in Table \ref{tab-be2a}. 

In Fig.~\ref{fig-be2-1} we present the intraband $B(E2)$ values for the yrast band. In panel (a) the experimental data are depicted while in panel (b) the theoretical ones are shown. We start by emphasizing that the $B(E2; 2_1^+\rightarrow 0_1^+)$ reduced transition probability is used to normalize the theoretical calculations and, consequently, these data  are well reproduced by the theoretical calculations. One notices that the increasing collectivity when approaching the midshell, although with its maximum at $A=104$, is correctly reproduced by the IBM-CM. At the theoretical side, one also observes an extra tiny peak  at $A=98$ superimposed to the up sloping trend for $B(E2; 6_1^+\rightarrow 4_1^+)$ and $B(E2; 8_1^+\rightarrow 6_1^+)$.

In Fig.~\ref{fig-be2-1} we depict two interband $B(E2)$ values involving the states $0_2^+$, $2_1^+$ and $2_2^+$,  in panel (a) for the experimental data while in panel (b) for the IBM-CM results. Considering the scarce available experimental information,  it is not possible to extract a well defined trend. At the theoretical side, conclusions are not obvious to be made although the agreement with the data is reasonable and two peaks are observed, one at $A=96$, corresponding to the $d_{5/2}$ subshell filling and the other at $A=104$, corresponding to reaching midshell. This behavior is pointing to a change in the structure of the states $0_2^+$, $2_1^+$ and $2_2^+$, which is readily confirmed with Fig.~\ref{fig-wf}. 

A more detailed comparison between experiment and theory is presented in Table \ref{tab-be2a} where all the available experimental information involving positive parity states is presented and compared with the IBM-CM results. The agreement between experiment and theory is noticeable and in most of cases  the theoretical values coincide with the experimental data within the error bar, although in several cases the error bar is too big. Note that the experimental $B(E2)$ allows to establish a stringent constraint on the Hamiltonian and transition operator parameters. As a matter of fact, the very large value of $\omega$ for $A=94$ is due to the rather large $B(E2)$ values for interband transitions.

\begin{figure}[hbt]
  \centering
  \includegraphics[width=1\linewidth]{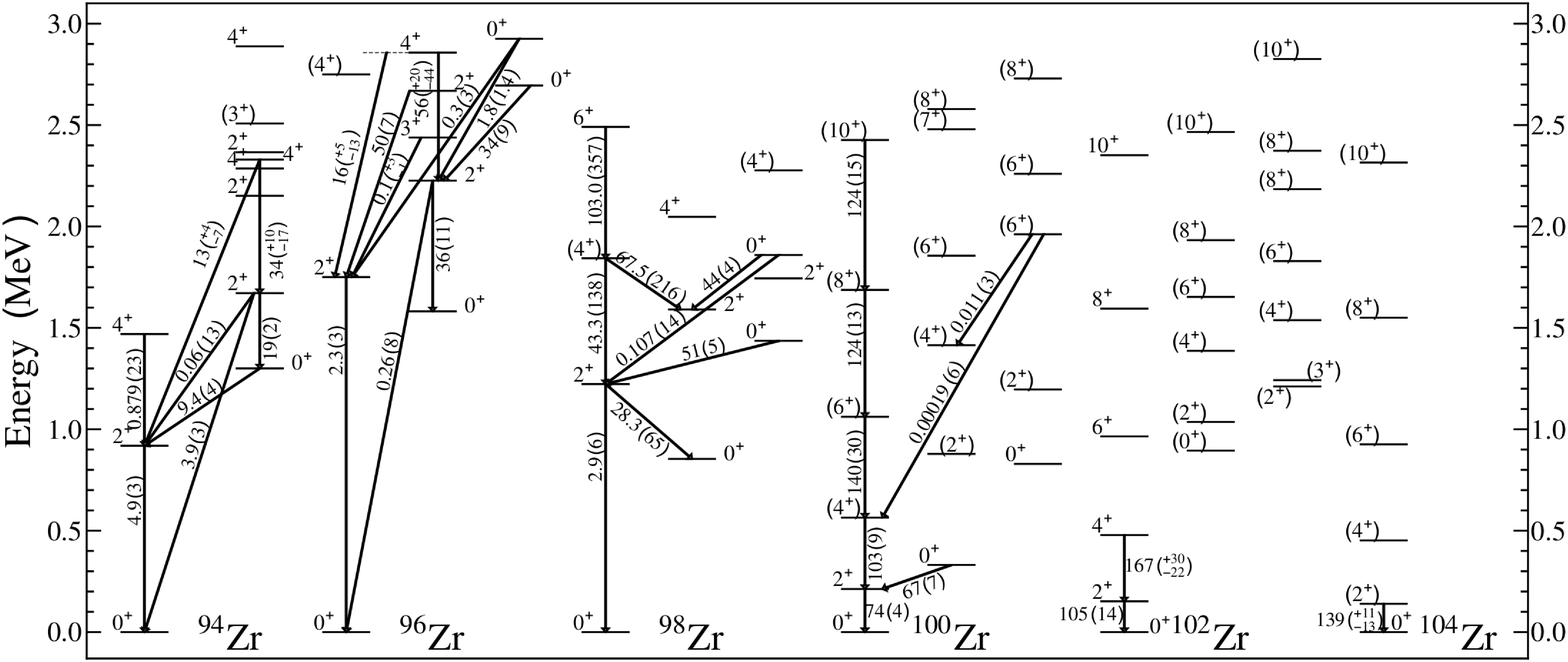}
  \caption{Experimental excitation energies and absolute $B(E2)$ transition rates for selected states in $^{94-104}$Zr.} 
  \label{fig-exp-96-104}
\end{figure}

\begin{figure}[hbt]
  \centering
  \includegraphics[width=1\linewidth]{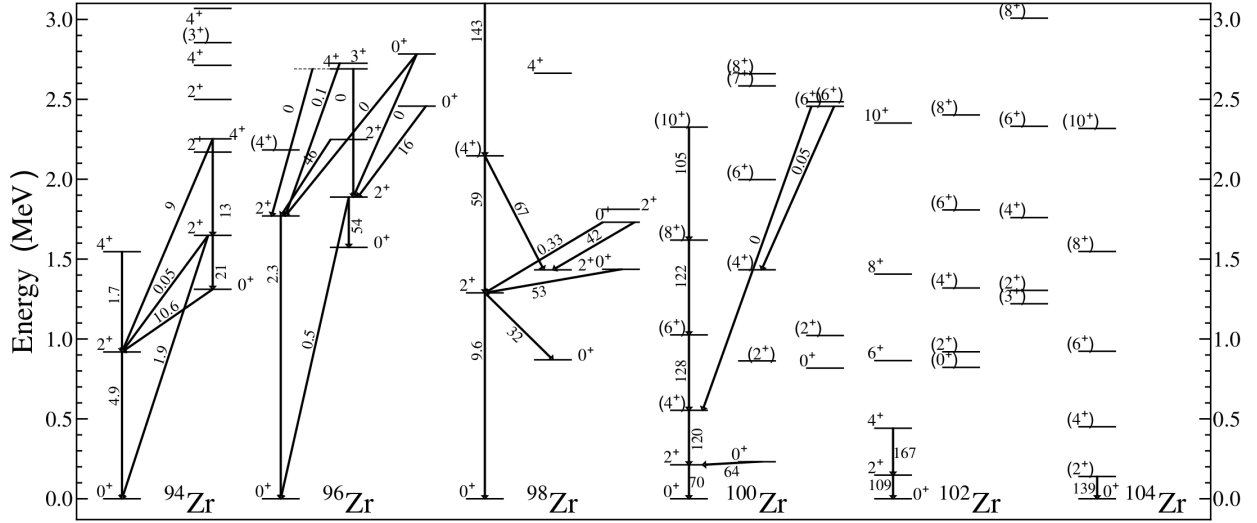}
  \caption{Theoretical excitation energies and absolute $B(E2)$ transition rates for selected states in $^{94-104}$Zr.} 
  \label{fig-theo-96-104}
\end{figure}

In Figs.~\ref{fig-exp-96-104} and \ref{fig-theo-96-104} we present the experimental and theoretical energy spectra (up to $E\approx 3$ MeV) and all known absolute $B(E2)$ values for masses $A=94-104$, which is the region where the coexistence should be more evident and the most rapid changes in energies and transition rates happen. The way of plotting the different states has been, first, in terms of an yrast band and, second, with a ``band'' built on a $0^+$ state that include one state per angular momentum. In the case that more states are known they have been grouped in an extra ``band''. The $0^+_1$ and  $0^+_2$ states own a different nature, either regular or intruder, but they could not share their nature with the rest of states of the ``band'', which means that states placed on different bands can be strongly connected by a large $B(E2)$ value, or the other way around, i.e., states on the same band could be weakly connected. 

In the case of $^{94}$Zr, the shown $B(E2)$ pattern indicates the presence of a rather large mixing. The $2_2^+$ state is tightly connected with the $0_2^+$ state, of mainly intruder nature, but, at the same time, it also presents a sizeable $B(E2)$ value into the ground state, which has a clear regular character. The same also holds for the $4_2^+$ state which is connected with two $2^+$ states of mixed character. In $^{96}$Zr the yrast band is barely connected with the intruder states, except the $2_1^+$ state that is connected with the $2_3^+$ and $4^+_2$ states, although in this latter case the experimental data presents a large uncertainty, compatible also with a rather small transition rate. On the other hand, the $2_2^+$ state is strongly connected with $0_2^+$, which points towards the intruder character of both states. In $^{98}$Zr, the $B(E2)$ indicate hints for the presence of a deformed band made of $0_2^+$, $2_1^+$, $4_1^+$, and $6_1^+$ states, while the ground state presents a regular character and it is hardly connected with the rest of states. The $0_3^+$ state is well connected with $2_1^+$ while $0_4^+$ with $2_2^+$. In the rest of isotopes $^{100-104}$Zr both the energies and the $B(E2)$ values point towards the presence of a well deformed and intruder yrast band and a set of non-yrast bands of regular character. Note that all the fine details discussed above are well described within the context of the IBM-CM calculations. 

\subsection{Wave function structure: from the unperturbed structure  to configuration mixing}
\label{sec-evolution}
\begin{figure}[hbt]
  \centering
  \includegraphics[width=.6\linewidth]{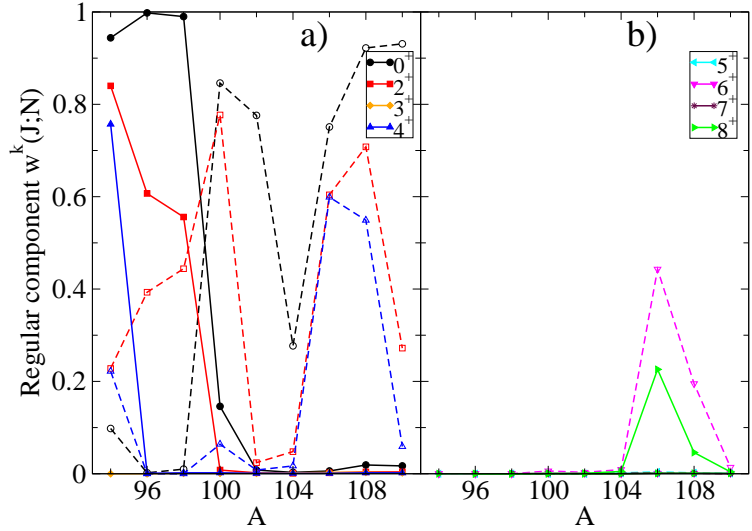}
  \caption{Regular content of the two lowest-lying states
    for each $J$ value (full lines with
    closed symbols correspond with the first state while dashed lines with open
    symbols correspond with the second state) resulting from the IBM-CM
    calculation, as presented in figure \ref{fig-energ-comp}.}
  \label{fig-wf}
\end{figure}

We start our analysis with the structure of the configuration-mixed wave functions along the first few levels, expressed as a function of the $[N]$ and $[N+2]$ basis states, as given in Eq.~(\ref{eq:wf:U5}). In Fig.~\ref{fig-wf}a and Fig.~\ref{fig-wf}b, we present 
the weight of the wave functions contained within the $[N]$-boson subspace, defined as the sum of the squared amplitudes $w^k(J,N) \equiv \sum_{i}\mid a^{k}_i(J;N)\mid ^2$, for both the yrast states, $k=1$, and the $k=2$ states (the latter are indicated with a dashed line) for spins $J^\pi=0^+, 2^+, 3^+, 4^+$ in panel (a) and $J=5^+, 6^+, 7^+, 8^+$ in panel (b).
In panel (a) one notices that the $0_1^+$ state has an almost pure regular character from $A=94$ up to $A=98$, while from $A=100$ and onwards the character turns to be purely intruder. The $0_2^+$ state starts being almost purely intruder for $A=94-98$, but then it becomes rather mixed, from $A=100$ and onwards. The $2_1^+$ state presents a clear trend, passing from being almost purely regular for $A=94$ into a purely intruder for $A=100$ and onwards. The $2_2^+$ state has an almost complementary trend with the $2_1^+$ state till $A=100$, but then becomes purely intruder for $A=102$ and $104$ and of mixed nature for $A=106-110$. The two $3^+$ states considered here present an intruder character for the whole isotope chain. In the case of the $4_1^+$ state, this state has a very large regular component for $A=94$, but is purely intruder from $A=96$ and onwards. All the presented $4^+$ states have an almost intruder character for the major part of isotopes, except for $A=106-108$, where a rather mixed character is shown. In the case of panel (b), the two considered states for $J^\pi=5^+, 6^+, 7^+,$ and $8^+$ show an intruder character for the whole chain, except the states $6_2^+$ and $8_1^+$, for $A=106$ and $108$ which present a certain degree of mixing, although keeping the intruder character.

\begin{figure}[hbt]
  \centering
  \includegraphics[width=.6\linewidth]{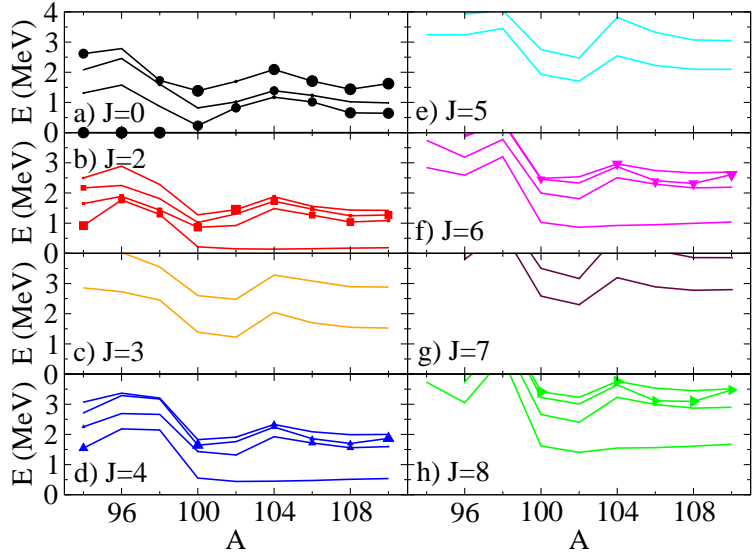}
  \caption{Energy systematics of the four lowest states below $4$ MeV. The size of the symbol is proportional to the value of $w^k(J,N)$ (see text for details). Each panel corresponds to a given angular momentum, (a) for $J=0$, (b) for $J=2$, (c) for $J=3$, (d) for $J=4$, (e) for $J=5$, (f) for $J=6$, (g) for $J=7$, and (h) for $J=8$.}
  \label{fig-wf-ener}
\end{figure}

In order to gain a deeper understanding on the systematics shown in Fig.~\ref{fig-wf} it will be of interest to present at the same time as the information on the wave function, the information on the excitation energy. This is shown in Fig.~\ref{fig-wf-ener} where we plot the energy systematics as a function of mass for the four lowest levels below $4$ MeV per angular momentum  and the size of the dot is proportional to the value of the regular component $w^k(J,N)$. In the case of panel (a), corresponding to $J=0$ one can appreciate how the main part of the regular component is moving from the ground state for $A=96-98$ to higher-lying states in $A=100$ and onwards. This can also be understood as the crossing and eventually mixing of several states with quite different character. For $J=2$ and $J=4$ (panels (b) and (d)), the major part of the regular component is distributed between several states, the first one for $A=94$, and the second, the third or the fourth for the rest, or even into higher-lying states (not shown), as for example, in the case of $J=4$ for $A=96$, $98$ and $102$. These results can be understood because of the crossing of different states relatively close in energy but with different nature. Finally, the cases $J=3, 5, 6, 7,$ and $8$ (panels (c), (e), (f), (g) and (h)), in the majority of cases correspond with states with an almost pure intruder character, being therefore the regular states well above the intruder ones.   

An additional  decomposition of the wave function is obtained by first calculating the wave functions within the N subspace as
\begin{equation}
  \Psi(l,JM)^{reg}_N = \sum_{i} c^{l}_i(J;N) \psi((sd)^{N}_{i};JM)~,
  \label{eq:wf:N}
\end{equation}
and likewise for the intruder (or N+2 subspace) as
\begin{equation}
\Psi(m,JM)^{int}_{N+2} = \sum_{j} c^{m}_j(J;N+2)\psi((sd)^{N+2}_{j};JM)~,
\label{eq:wf:N+2}
\end{equation}
defining an ``intermediate'' basis \cite{helle05,helle08} that corresponds to the states appearing in Fig.\ \ref{fig-ener-nomix}, where the mixing term, $\omega$, was cancelled. This generates a set of bands within the 0p-0h and 2p-2h subspaces, corresponding to the unperturbed bands that are extracted in schematic two-level phenomenological model calculations  (as discussed in references \cite{bree14,liam14,kasia15,duppen90,drac88,drac94,allatt98,page03,drac04}).
\begin{figure}
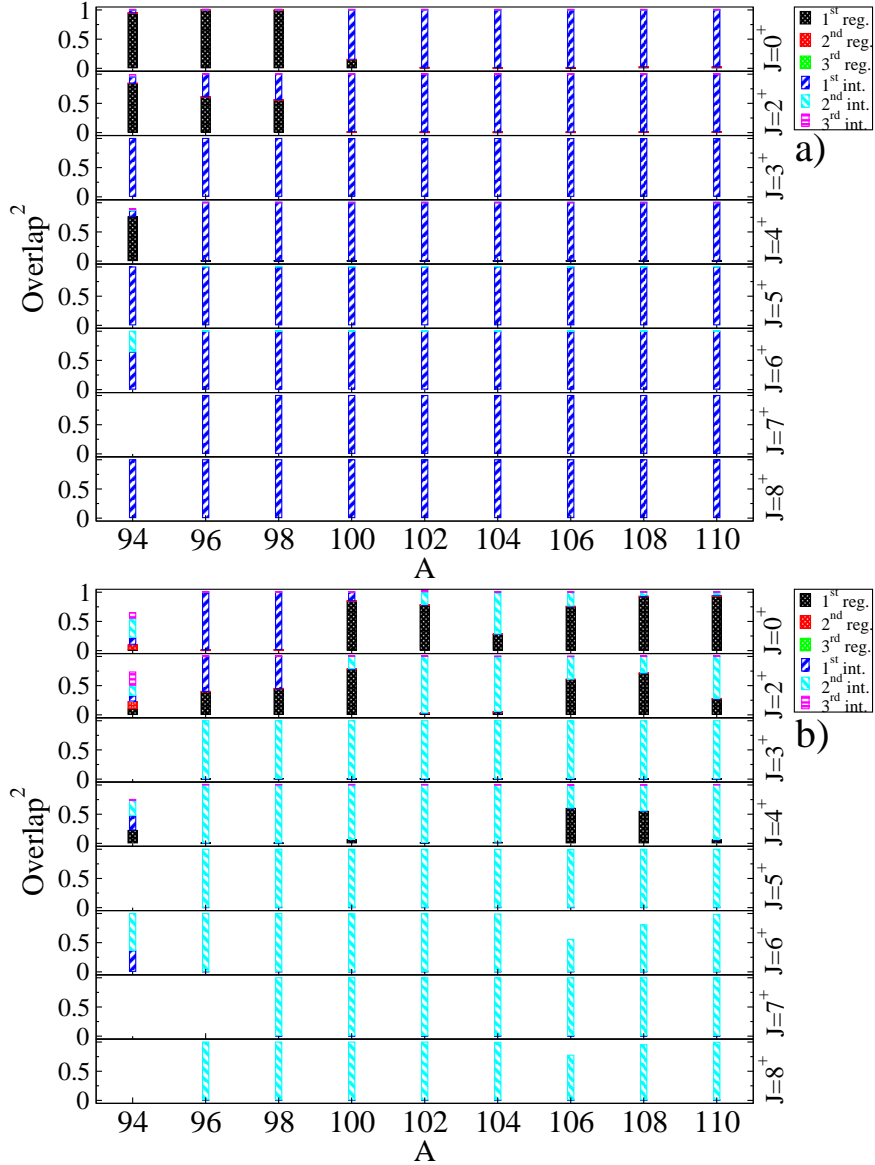

\includegraphics[width=.7\textwidth]{overlap-int-reg-1.eps}\\
\includegraphics[width=.7\textwidth]{overlap-int-reg-2.eps}
\caption{Overlap of the wave functions of Eq.~(\ref{eq:wf:U5}), with the wave functions describing the unperturbed basis  Eq.~(\ref{eq:wf:N}) and Eq.~(\ref{eq:wf:N+2}). Panel (a): overlaps for first $0^+,2^+,3^+,4^+;5^+,6^+,7^+,8^+$ state, panel (b): overlaps for the corresponding second state (see also text).}
\label{fig-overlap} 
\end{figure} 

The overlaps $_{N}\langle l,JM \mid k,JM\rangle$ and $_{N+2}\langle m,JM \mid k,JM\rangle$ can then be expressed as,
\begin{equation}
_{N}\langle l,JM \mid k,JM\rangle=\sum_{i} a^{k}_i(J;N) c^{l}_i(J;N), 
\end{equation} 
and  
\begin{equation}
_{N+2}\langle m,JM \mid k,JM\rangle=\sum_{j} b^{k}_j(J;N+2) c^{m}_j(J;N+2),
\end{equation} 
(see expressions (\ref{eq:wf:N}) and (\ref{eq:wf:N+2})).

In Fig.~\ref{fig-overlap} we show these overlaps, but squared, where we restrict ourselves to the first and second state ($k=1,2$) with angular momentum J$^{\pi}= 0^+, 2^+, 3^+, 4^+, 5^+, 6^+, 7^+, 8^+$, and give the overlaps with the lowest three bands within the regular $(N)$ and intruder $(N+2)$ spaces ($l=1,2,3$ and $m=1,2,3$).  Since these figures are given as a function of mass number, one obtains a graphical insight into the changing wave function content. In the case of the first state, $k=1$, the $0^+$ states show up a sudden transition from the first regular component ($A=94-98$) to the first intruder one ($A=100-110$); the $2^+$ states present a large overlap with the first regular component in $A=94$, there is an almost $50\%$ mixing of the first regular and the first intruder component in $A=96$ and $98$, and it moves  into a large first intruder component in $A=100$ and onwards; the $4^+$ state corresponds to the first regular component in $A=94$, but it presents a complete overlap with the first intruder state in the rest of the isotope chain; the rest of angular momenta always present a large overlap with the first intruder state except the $6^+$ state in $A=94$ were the overlap is with the first and the second intruder state.
Passing to the second states ($k=2$), the $0^+$ state is mainly of intruder character for $A=94$, $96$ and $98$, with  a combination of several intruder components at $A=94$, whereas at $A=96-98$ a large overlap with the second intruder state results. In the nucleus with $A=100$, there is a strong overlap with the first regular state. The region $A=102-110$ corresponds to an overlap with the first regular and the second intruder states, but with and increasing overlap with the regular state as A increases. Here, the $2^+$ states show a very similar qualitative behavior to $0^+$, but with a full overlap with the second intruder state for $A=102$ and $104$; the $4^+$ state corresponds in $A=94$ to a mixture between the first regular and the first and second intruder states, in the case of $A=96-104$ and $110$ the overlap is complete with the second intruder state, while for $A=106-108$ presents a $50\%$ mixing between the first regular and the second intruder state; the rest of angular momenta all correspond to a large overlap with the second intruder state, except $J^+=6^+$ in $A=94$ which presents a mixture between the first and the second intruder states.

The lack of overlap for some masses and angular momenta is due to the small size of the basis in theses cases, for example the dimension of the Hilbert space for $J^+=3^+$ in $A=94$ is $1$ and this state sits in the intruder sector, which is the reason why the overlap with the first intruder state is $100\%$ in the case of $k=1$ and no overlap is shown for $k=2$.  

Summarizing, Fig.~\ref{fig-overlap} shows the predominance of the intruder sector in low-lying states, except for $J^+=0^+$ and $2^+$ where a large mixing appears and for $4^+$, where there is certain overlap with the first regular state, but the main overlap is with the first intruder one.

\section{Study of other observables: radii, isotopic shifts, E0 transition rates and two-neutron separation energies}
\label{sec-other}

\begin{figure}[hbt]
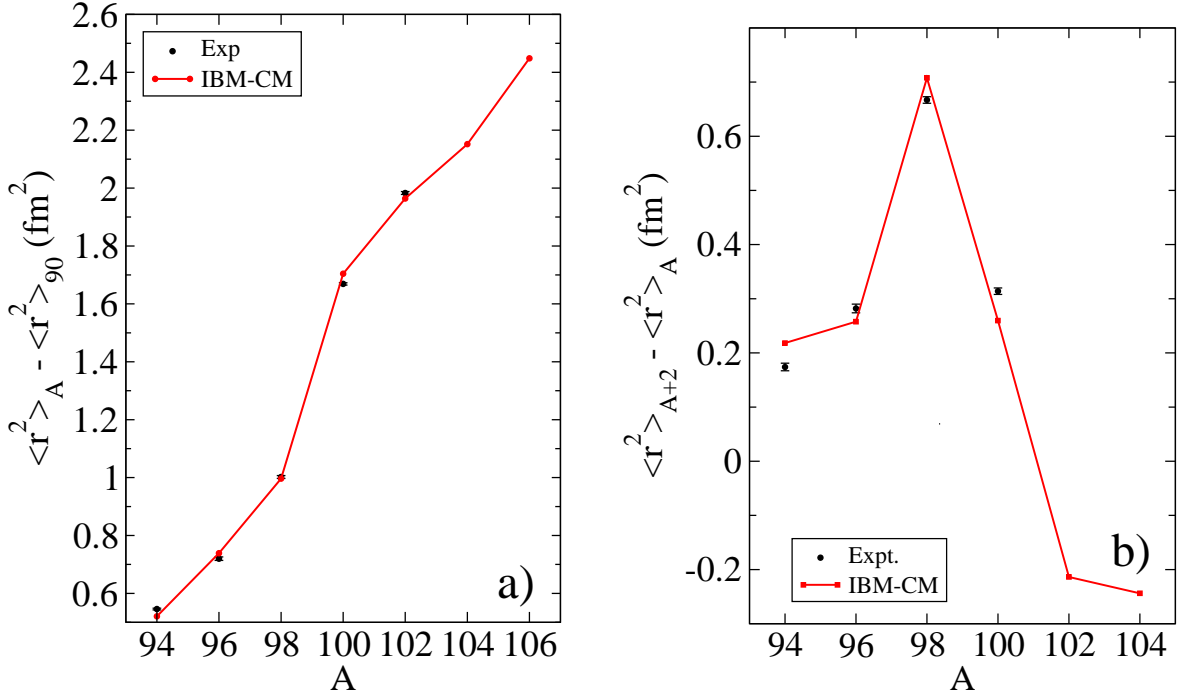
 
  \centering
  \includegraphics[width=.45\linewidth]{radius.eps}
~~~~
  \includegraphics[width=.45\linewidth]{iso-shift.eps}
  \caption{ (a) Charge mean-square radii for the Zr nuclei. (b) Isotopic shift for the Zr nuclei. The data are taken from the column corresponding to optical isotope shifts of \cite{Ange13}.
  }
  \label{iso-shift}
\end{figure}

\subsection{Radii, isotopic shifts and E0 transitions}
\label{sec-radii}
Nuclear radii provide direct information on the evolution of nuclear deformation, as shown in Fig.\ \ref{fig-s2n-radii}a, where one observes a sudden increase of the nuclear radius when passing from $^{99}$Zr to  $^{100}$Zr, as well as in the rest of nuclei of this mass region at $N=60$. Experimental information on radii, for both even-even and even-odd nuclei, can be found in \cite{Ange13} and in this section we will confront the IBM-CM predictions with the experimental data.

To calculate the nuclear radius, we have used the standard IBM-CM expression, 
\begin{equation}
r^2=r_c^2+ \hat{P}^{\dag}_{N}(\gamma_N \hat{N}+ \beta_N
\hat{n}_d)\hat{P}_{N} + 
\hat{P}^{\dag}_{N+2}(\gamma_{N+2} \hat{N}+ \beta_{N+2} \hat{n}_d) \hat{P}_{N+2}.
\label{ibm-r2}
\end{equation} 
The four parameters appearing in this expression have been fitted to the experimental data, corresponding to the radii of the $A=94-106$ even-even Zr isotopes \cite{Ange13}, starting from the Hamiltonian parameters already extracted and discussed in Section \ref{sec-fit-procedure} (see Table \ref{tab-fit-par-mix}). We only consider nuclei in the first part of the nuclear shell $N=50-82$, namely, up to $A=106$, on one hand, because experimental data are only known till $A=104$ (see \cite{Ange13}) and, on the other hand, because the parameters appearing in Eq.\ (\ref{ibm-r2}) could change when crossing midshell, in a similar way to what happens for the case of S$_{2n}$ (see Ref.\ \cite{Foss02} for details). Moreover, if they are assumed to be equal, the following transformation holds, $\beta_N^{(2)}=\beta_N^{(1)}$,  $\beta_{N+2}^{(2)}=\beta_{N+2}^{(1)}$, $\gamma_N^{(2)}=-\gamma_N^{(1)}$, $\gamma_{N+2}^{(2)}=-\gamma_{N+2}^{(1)}$, where the superindex $(1)$ refers to the first half part of the shell while $(2)$ to the second. Besides, in order to avoid a discontinuity when crossing the midshell, it is needed to add to the value of $r^2$, in the second half of the shell, the term $\gamma^{(1)}_N N_0 \omega +\gamma^{(1)}_{N+2}(N_0+4)(1-\omega)$, where $N_0$ is the number of bosons at the midshell and $w=w^1(0,N)$ (see Eq.\ (\ref{wk})) the regular content of the ground state. The resulting values are  $\gamma_N=0.245$ fm$^2$, $\beta_N=0.195$ fm$^2$,  $\gamma_{N+2}=0.275$ fm$^2$, and  $\beta_{N+2}=-0.09$ fm$^2$. In determining these parameters, we took as reference point the value of the radius of  $^{90}$Zr.

\begin{table}
\caption{Comparison of experimental \cite{Wood99,Kibe05} and theoretical values of $\rho^2(E0)$ values given in miliunits.}
\label{tab-e0}
\begin{tabular}{llcc}
  \hline
  \hline
  Isotope & Transition & $10^3 \rho^2(E0)$ (Exp) & $10^3 \rho^2(E0)$ (Theo)\\
  \hline
  $^{94}$Zr & $0_2^+\rightarrow 0_1^+$& 11.5 (11) & 8.9 \\
  $^{96}$Zr & $0_2^+\rightarrow 0_1^+$& 7.57 (14) & 3.7 \\
  $^{98}$Zr & $0_2^+\rightarrow 0_1^+$& 11.2 (12) & 21 \\
  $^{98}$Zr & $0_3^+\rightarrow 0_2^+$& 76 (6) & 13 \\
  $^{98}$Zr & $0_4^+\rightarrow 0_3^+$& 61 (8) & 7.4 \\
  $^{100}$Zr & $0_2^+\rightarrow 0_1^+$& 108 (19) & 150 \\
  \hline
  \hline
\end{tabular}
\end{table}

In panel (a) of Fig.~\ref{iso-shift}, the value of the charge mean-square radius is depicted. One notices the rapid increase of the radius in moving from $A=98$ to $A=100$ ($N=60$). The change is more clearly observed in panel (b) of this figure, in which the isotopic shift is depicted. In particular, its value starts from $\approx$ $0.3$ fm$^2$ in $^{96}$Zr to $\approx$ $0.7$ fm$^2$ in $^{98}$Zr, decreasing to $0.3$ fm$^2$ in $^{100}$Zr, pointing out  a sudden change of deformation in $^{100}$Zr. To have an idea of how big this change is, it is worth mentioning  that in the well-known region of lead, where shape coexistence plays a major role, in the Pt nuclei, the maximum change of the isotopic shift is $\approx$ $0.07$ fm$^2$, in Hg $\approx$ $0.02$ fm$^2$, while in Po $\approx$ $0.05$ fm$^2$. The agreement between theory and experiments is satisfactory in spite of the small error bars. This points towards an appropriate description of the  ground-state wave functions, in particular, regarding the balance between regular and intruder sectors, but also regarding the deformation.  We can also provide the value of the radius for the two isotopes, $A=104$ and $106$, for which no experimental information concerning radii is available. The value of $r^2$ (relative to $^{90}$Zr) is $2.15$ fm$^2$ and $2.45$ fm$^2$ for $^{104}$Zr and $^{106}$Zr, respectively.

Once the parameters appearing in (\ref{ibm-r2}) have been fixed, it is also possible to calculate the  $\rho^2$(E0) transition rates, as shown in \cite{Zerg12}. In particular, the transition operator can be defined as,
\begin{equation}
T(E0)=(e_n N+ e_p Z) r^2  
\end{equation}
where we take as effective charges, the standard ones, namely, $e_n=0.5$e and $e_p=1$e \cite{Wood99}, being $e$ the proton charge. It is customary to use the derived quantity $\rho^2 (E0)$, defined as,
\begin{equation}
  \label{eq:E0}
  \rho^2(E0)_{J_i\rightarrow J_j}=\left(\frac{\left\langle J_i| r^2|J_j\right\rangle}{e R^2}\right)^2,
\end{equation}
where $R=1.2 A^{1/3}$ fm. Moreover, we emphasize the experimental variation in the value of $\rho^2(E0)$, which is directly connected with the isomeric shift, $\Delta \langle r^2 \rangle$, between the first excited state, 0$^+_2$, and the ground-state, 0$^+_1$.  We refer to the study by \citet{Zerg12} who also carried out IBM calculations, however not including configuration mixing. This implies that there is no contribution from the $\hat{N}$ term of Eq.~(\ref{ibm-r2}) to the value of $\rho^2$(E0). But, when considering configuration mixing, the $\hat{N}$ term generates a non-vanish contribution from the mixing between regular and intruder sectors in the wave function. In Table \ref{tab-e0}, we present the theoretical results that are compared with the available experimental information, taken from \cite{Wood99,Kibe05}. The data on $\rho^2(E0)$ ($0^+_2 \rightarrow 0^+_1$) exhibit a strong increase from $^{90}$Zr onwards, reaching the value of $108(19)$ miliunits in $^{100}$Zr. This pattern is illustrated in Figs.\ 9 and 10 of Ref.\ \cite{Wood99}. This is given a hint for a strong change in the underlying structure of the 0$^+_{1,2}$ states in the Zr isotopes when moving from $N=50$ towards $N=60$.
We observe a rather good agreement with the experimental results concerning the transition $0_2^+ \rightarrow 0_1^+$. However, when other high-lying $0^+$ states are involved in the transition, such as $0_3^+$ and $0_4^+$, the agreement is no longer so good. In particular, the predicted values for $^{98}$Zr are too small, which points towards a bad description of the particular states  $0_3^+$ and $0_4^+$.

\subsection{Two-neutron separation energies}
\label{sec-s2n}
Two-neutron separation energies, $S_{2n}$, have been traditionally  used as an indicator of the existence of Quantum Phase Transitions (QPTs) \cite{Cejn10} and, in general, they also serve as a hint for the onset of deformation. Indeed, in Ref.~\cite{Garc05}, the chain of Zr isotopes was studied, using the IBM with a single configuration, paying special attention to the systematics of $S_{2n}$ as an indicator of the appearance of a QPT at $A\approx 100$, but also of the onset of deformation.
\begin{figure}[hbt]
\centering
\includegraphics[width=0.50\linewidth]{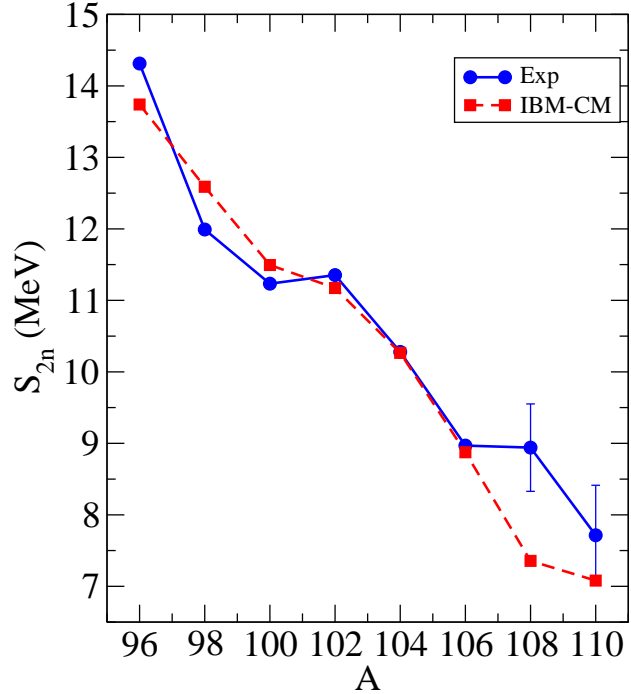}
\caption{Comparison of experimental and theoretical two-neutron separation energies.}
\label{fig-s2n-zr}
\end{figure}

The definition of S$_{2n}$ involves the binding energy of two neighbouring nuclei separated by two units of mass, as written in this equation,
\begin{equation}
\label{s2n}
S_{2n}(A)=BE(A)-BE(A-2). 
\end{equation}
where $BE$ denotes the binding energy of the nucleus. Note that in order to calculate the value of S$_{2n}$, one has to add to the binding energy extracted considering the IBM-CM, a linear trend due to the bulk (or global) contribution from the Hamiltonian that does not affect the energy spectrum. This term results from a Liquid Drop Model contribution. In the case of a single configuration it is customary to assume the expression,
\begin{equation}
S_{2n}(N)={\cal A} +{\cal B} N+BE^{lo}(N)-BE^{lo}(N-1),
\label{s2n-lin}
\end{equation}
where $BE^{lo}$ is the local binding energy derived from the IBM Hamiltonian, and the coefficients ${\cal A}$ and ${\cal B}$ come from the bulk contribution. Although the later expression can also be used under the presence of intruder states, we propose to use, as an \textit{ansazt}, 
\begin{equation}
S_{2n}(N)={\cal A} +{\cal B} (N+2(1-w))+BE^{lo}(N)-BE^{lo}(N-1),
\label{s2n-lin-new}
\end{equation}
where $w=w^1(0,N)$ (see Eq.\ (\ref{wk})), i.e., the regular content of the ground state, and the value of ${\cal B}$ has been assumed to be equal in the regular and in the intruder sector. This expression takes into account the effect of the intruder sector to the bulk contribution, adding two extra bosons weighted with the intruder content of the ground state. These coefficients should be obtained from a least-squares fit to the experimental S$_{2n}$ value, assuming they are constant along the whole chain of isotopes (see \cite{Foss02} for the details of the method). Of course, Eqs.\ (\ref{s2n-lin}) and (\ref{s2n-lin-new}) can also be  written in terms on the atomic mass $A$.
Indeed, the fitted values of ${\cal A}$ and ${\cal B}$ when using the atomic mass correspond to ${\cal A}=30.7$ MeV and ${\cal B}=-0.377$ MeV. For identical reason than in Section \ref{sec-radii}, the fit has been restricted to the first half part of the shell, although in this case experimental data exist, but with a large error bar. Assuming there is no change in ${\cal A}$  and ${\cal B}$ when crossing the midshell, the same values can be used in the whole shell when using Eqs.\ (\ref{s2n-lin}) or (\ref{s2n-lin-new}) written in terms of the atomic mass. If instead, the boson number, $N$, is used, the sign of ${\cal B}$ will change and an offset should be added.

This observable is very sensitive to the rapid changes of nuclear structure, because, as it was explained in Section \ref{sec-corr_energy}, the binding energy is highly dependent on the deformation. In Fig.\ \ref{fig-s2n-zr}, experimental and theoretical values of S$_{2n}$ are depicted, obtaining a reasonable agreement between the experimental data and the theoretical calculations. In this figure one can readily see a change in the slope of S$_{2n}$ when approaching $A\approx 100$, which is approximately reproduced by the theoretical results. These results are very close to the values obtained using Eq.\ (\ref{s2n-lin}) to extract the global contribution. The systematics hints the rapid onset of deformation, but it is not possible to distinguish if the reason is the crossing of two families of states with different deformation or is the growing of deformation in a single family of states. Note that the calculations have been extended to the second half part of the shell, where the agreement is not good, but it should be taken into account that still the experimental data in this area are not reliable enough and they have a large error bar.

\begin{figure}
\includegraphics[width=0.3\textwidth]{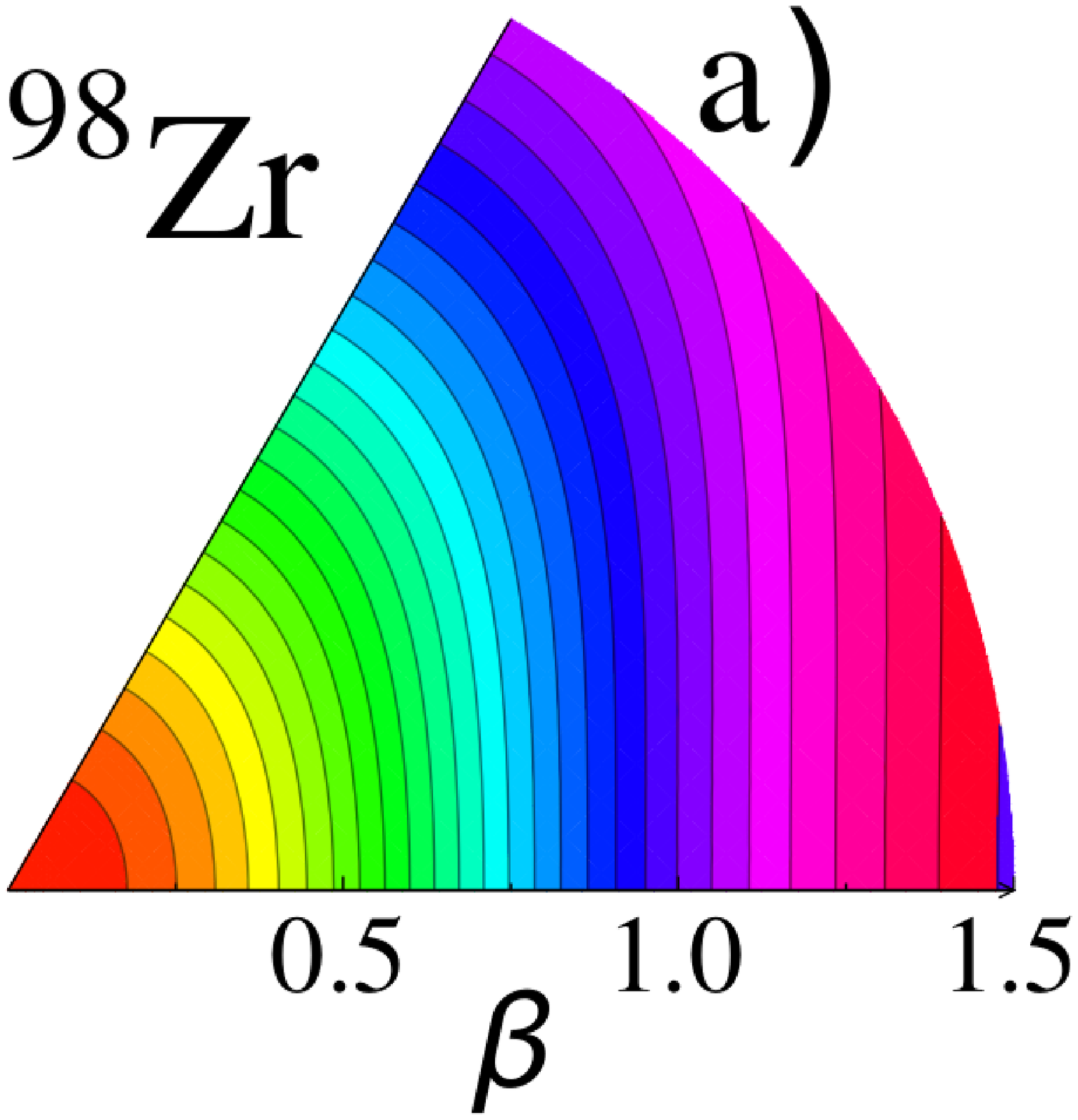}%
\includegraphics[width=0.3\textwidth]{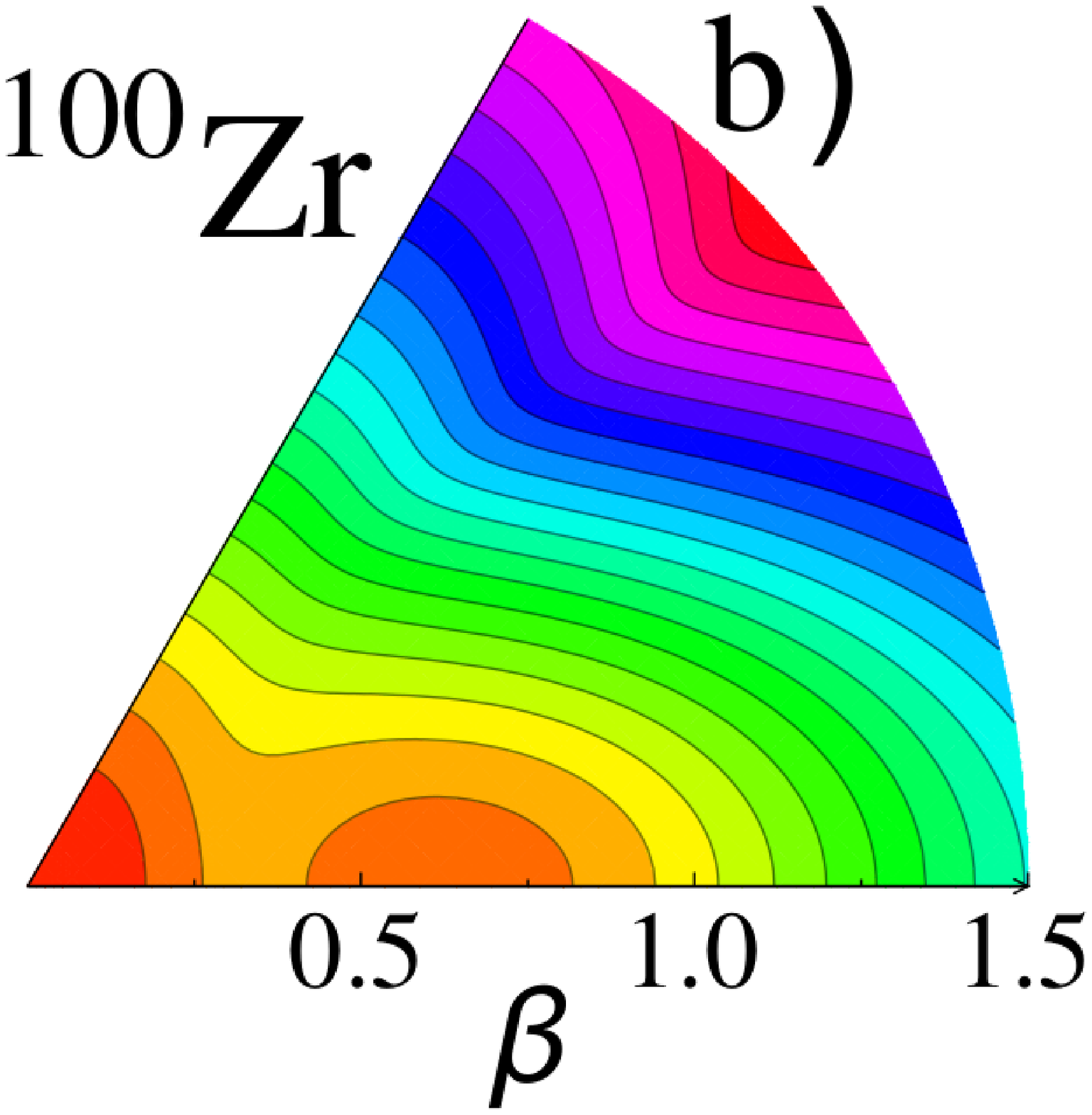}%
\includegraphics[width=0.3\textwidth]{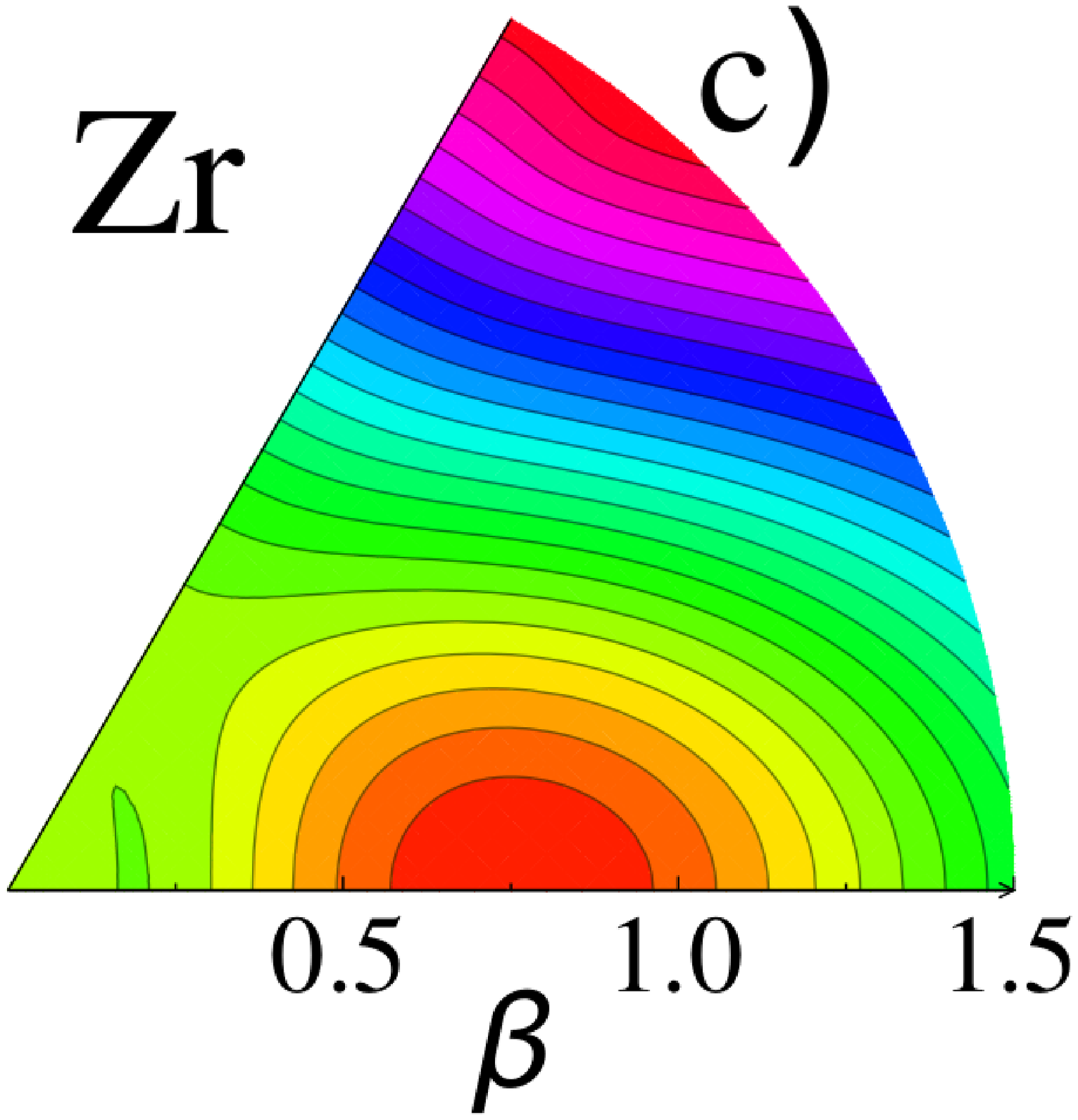}
\caption{Matrix coherent-state calculation for $^{98}$Zr in panel (a), $^{100}$Zr in panel (b), and $^{102}$Zr in panel (c), corresponding with the present IBM-CM Hamiltonian (table \ref{tab-fit-par-mix}). The energy spacing  between adjacent contour lines equals $100$ keV and the deepest energy minimum is set to zero, corresponding to the red color.}
\label{fig_ibm_ener_surph}
\end{figure}

\section{Quadrupole shape invariants, nuclear deformation and mean-field energy surfaces}
\label{sec-q-invariants}
The IBM is a model capable to provide in a straightforward way energy spectra and the corresponding wave functions, $B(E2)$ values and radii. We discuss in this Section how the model is also able to provide an interesting and precise insight on nuclear deformation and associated nuclear shapes. A model description, acting within a \textit{laboratory frame} does not contain the observable ``deformation'', however, it is possible to calculate certain observables that can be measured experimentally and that makes it possible to extract direct information on ``deformation''. This concept is usually defined within an intrinsic frame and derived with mean-field methods. Therefore, we also calculate the mean-field energy surface deformation corresponding with the algebraic IBM-CM model approach.
\begin{table}
\caption{Values of deformation, $q^2$, and $\gamma$, extracted from quadrupole shape invariants, together with the value of $w^k$ ([N] content as defined in Eq.\ (\ref{wk})).}
\label{tab-q-invariant}
  \begin{center}
\begin{tabular}{ccccc|ccccc}
\hline
A & State & q$^{\text{2}}$ (e$^{\text{2}}$ b$^{\text{2}}$) & ~~$\gamma$~~ & $w^k$&
~~A~~ & State & q$^{\text{2}}$ (e$^{\text{2}}$ b$^{\text{2}}$) & ~~$\gamma$~~ & $w^k$\\
\hline
94 & 0$_{\text{1}}^+$ & 0.11 & 49 & 0.944& 104 & 0$_{\text{1}}^+$ & 2.04 & 2 & 0.003\\   
  & 0$_{\text{2}}^+$ & 0.31 & 43 & 0.098&    & 0$_{\text{2}}^+$ & 0.97 & 0 & 0.277  \\      
  & 0$_{\text{3}}^+$ & 0.22 & 32 & 0.034&    & 0$_{\text{3}}^+$ & 0.41 & 0 & 0.719  \\    
  \hline                                      
96 & 0$_{\text{1}}^+$ & 0.04 & 52 & 0.998& 106 & 0$_{\text{1}}$ & 1.60 & 9 & 0.006  \\     
  & 0$_{\text{2}}^+$ & 1.16 & 31 & 0.002&    & 0$_{\text{2}}^+$ & 0.34 & 0 & 0.751  \\      
  & 0$_{\text{3}}^+$ & 0.66 & 35 & 0.003&    & 0$_{\text{3}}^+$ & 0.83 & 13 & 0.256 \\     
  \hline                                      
98 & 0$_{\text{1}}^+$ & 0.16 & 52 & 0.990& 108 & 0$_{\text{1}}^+$ & 1.27 & 10 & 0.019 \\   
  & 0$_{\text{2}}^+$ & 1.07 & 37 & 0.010&    & 0$_{\text{2}}^+$ & 0.14 & 0 & 0.922  \\      
  & 0$_{\text{3}}^+$ & 0.88 & 53 & 0.296&    & 0$_{\text{3}}^+$ & 0.83 & 15 & 0.068 \\     
\hline                                      
100 & 0$_{\text{1}}^+$ & 1.00 & 10 & 0.146&110 & 0$_{\text{1}}^+$ & 1.06 & 11 & 0.017 \\   
  & 0$_{\text{2}}^+$ & 0.22 & 0 & 0.846&      & 0$_{\text{2}}^+$ & 0.10 & 0 & 0.931 \\       
  & 0$_{\text{3}}^+$ & 0.56 & 14 & 0.012&     & 0$_{\text{3}}^+$ & 0.70 & 15 & 0.054 \\     
\hline
102 & 0$_{\text{1}}^+$ & 1.60 & 13 & 0.008 & & & & & \\
  & 0$_{\text{2}}^+$ & 0.29 & 0 & 0.776&  & & & & \\
  & 0$_{\text{3}}^+$ & 0.87 & 20 & 0.215& & & & & \\
\hline
\end{tabular}
\end{center}
\end{table}

It is possible to define a mean-field energy surface at the IBM-CM level, obtaining the corresponding energy in a $\beta-\gamma$ plane. This geometric interpretation of the IBM is built thanks to the so called intrinsic state formalism, as proposed in the seminal works \cite{gino80,diep80a,diep80b,Gilm74}. The extension of this formalism to describe regular and intruder configurations, i.e., the IBM-CM, was proposed by Frank {\it et al.}\, introducing a matrix coherent-state method \cite{Frank02,Frank04,Frank06,Mora08}, allowing to calculate the total energy surface that corresponds to the lowest eigenvalue of a two-by-two matrix. In Refs.~\cite{Garc14a,Garc14b,Garc15}, a detailed description of the method and its application to Pt, Hg, and Po isotopes, respectively, has been given.  In Fig.~\ref{fig_ibm_ener_surph}, the IBM-CM mean-field energy surfaces are presented for the isotopes $^{98}$Zr, $^{100}$Zr, and $^{102}$Zr in panels (a), (b), and (c), respectively. In these three nuclei, a rapid transition from a spherical to well-deformed shapes is observed. In panel (a) of Fig.~\ref{fig_ibm_ener_surph}, the energy surface of  $^{98}$Zr is depicted, presenting a well defined spherical minimum. In panel (b), $^{100}$Zr  exhibits the coexistence of two minima, one spherical and the other prolate deformed, that are almost degenerate. In panel (c), the energy surface of $^{102}$Zr presents a deep deformed minimum, but also a rather shallow spherical one. An almost similar situation happens for $^{104-110}$Zr (not shown). Note that according to the mean-field approach, $^{100}$Zr still has as deepest minimum the spherical one, although very close-lying in energy to the prolate one. In Refs.\ \cite{Rodr10}, \cite{Nomu16} and \cite{Mei12} these nuclei have also been studied, among many other, using a D1S Gogny interaction in the first case, using a Gogny D1M interaction and a mapping procedure into the IBM-CM Hamiltonian, in the second, and using a covariant density functional theory with the parameter set PC-PK1, in the third. In all cases, a prolate oblate coexistence is observed. 

A major question on how to obtain information on nuclear deformation within the laboratory frame was first addressed by \citet{kumar72} and \citet{Cline86}, introducing the concept of higher moments of the quadrupole operator (quadratic and cubic), a method that has been used extensively to extract the intrinsic $\beta,\gamma$ values as extra labels to be added to each nuclear eigenstate. From the experimental side, the extraction of the quadrupole $E2$ diagonal and non-diagonal matrix elements as well as their relative sign, using Coulomb excitation, has been of major importance leading to information on deformation in an almost model-independent way.  More recently, work carried out by Alhassid, Gilbreth and Bertsch have addressed the question of nuclear deformation within the laboratory  frame, even extending towards finite temperature (excited states) making use of the auxiliary-field Monte Carlo method \cite{Alha14,Gilb18}, an approach that was developed following the early work by \citet{koon97}.

Using experimental data, it is possible to extract direct information about various higher moments characterizing the nuclear shape corresponding with a given eigenstate. One of the most powerful techniques to this end is Coulomb excitation, which makes it possible to extract both the diagonal as well as the non-diagonal quadrupole and octupole matrix elements, including their relative signs and, in an almost  model independent way, extract information about nuclear deformation as shown by Kumar, Cline and co-workers \cite{kumar72,Cline86,Wu96,clement07,sreb11}. This method has also been used within the context of the IBM approach in \cite{Wern00}.

The key point of this approach is the use of the concept of an ``equivalent ellipsoid'' view of a given nucleus \cite{kumar72} that corresponds to a uniformly charged ellipsoid with the same charge, $\langle r^2 \rangle$, and  $E2$ moments as a specific eigenstate of a given nucleus \cite{kumar72,wood07}. 

To obtain the theoretical value of the nuclear deformation of a given state we will resort to the so-called quadratic and cubic quadrupole shape invariants. The quadrupole deformation for $0^+$ states corresponds to
\begin{eqnarray} 
q_{2,i}&=&\sqrt{5} \langle 0_i^+| [\hat{Q} \times \hat{Q} ]^{(0)}|0_i^+
\rangle,\\
q_{3,i}&=&-\sqrt{\frac{35}{2}} \langle 0_i^+[ \hat{Q} \times \hat{Q} \times \hat{Q}]^{(0)}|0_i^+
\rangle.
\label{q_invariant1}
\end{eqnarray} 
For the triaxial rigid rotor model \cite{wood09}, it is directly connected with the deformation parameters
\begin{eqnarray} 
q_2 &=& q^2\label{q_invariant2b},\\
q_3 &=& q^3 \cos{3~\delta}, 
\label{q_invariant2}
\end{eqnarray} 
where $q$ denotes the nuclear intrinsic quadrupole moment and $\delta$ the triaxial degree of freedom,
\begin{eqnarray} 
q&=&\sqrt{ q_2},\\
\delta &=&\frac{60}{\pi} \arccos{\frac{q_3}{q_2^{3/2}}}.
\label{q_invariant3}
\end{eqnarray} 
The value of $\delta$ is equivalent, up to first order  approximation, to the value of $\gamma$ appearing in the Bohr-Mottelson model (see Ref.~\cite{Srebrny06} for further details).

In Table \ref{tab-q-invariant}, we present the theoretical value of $q^2$ and $\gamma$, together with the fraction of the wave function belonging to the regular sector, $w^k$ (\ref{wk}), corresponding to the $0_1^+$, $0_2^+$  and $0_3^+$ states, for the whole chain of Zr isotopes. This table shows how very different shapes coexist, indeed, more than two, although only two different type of  configurations are used in our approach, 0p-0h and 2p-2h. In $^{94}$Zr, the three $0^+$ states are almost spherical, in $^{96-98}$Zr, the ground state remains spherical while the other excited states start to become deformed and approximately triaxial. In $^{100-104}$Zr, the ground state is well deformed and prolate, while the first regular excited state is slightly deformed and prolate, with a maximum of deformation at $A=104$, and the first intruder excited state is much less deformed than the ground state, but also of prolate nature. In summary, from $A=100$ and onwards, the prolate shapes dominate, while from $A=94$ till $A=98$, less deformed shapes, almost spherical in some cases, but of oblate nature, are the most abundant ones. Moreover, one can notice that the most deformed state in each isotope is always of intruder nature, while the most spherical one belongs to the regular sector.

\begin{figure}[hbt]
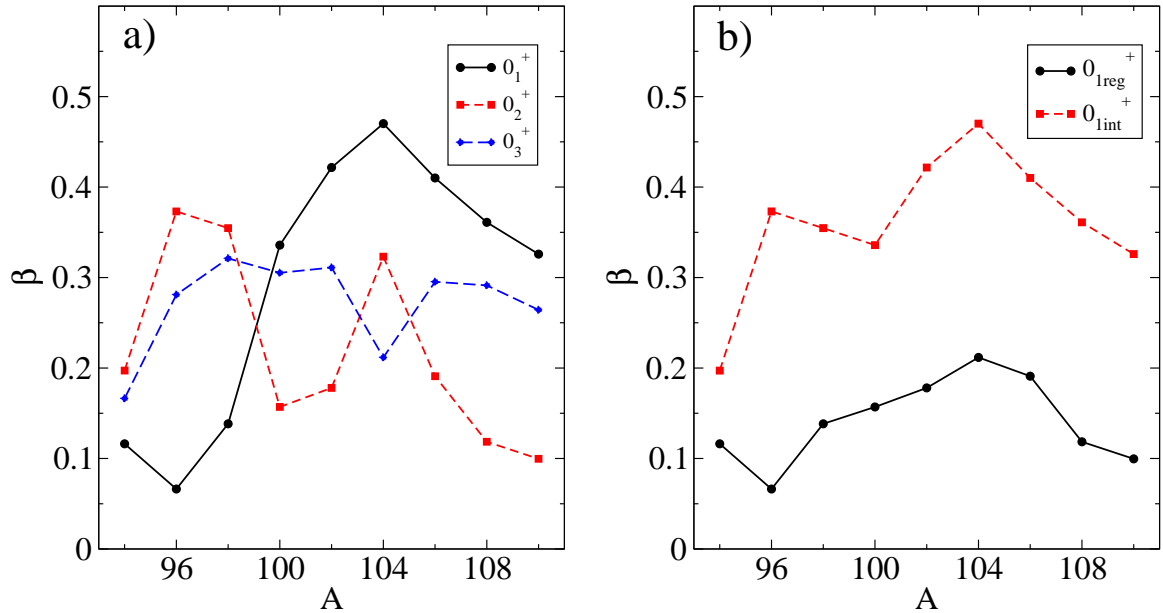

  \centering
  \includegraphics[width=.45\linewidth]{beta_fromQ2.eps}
  ~~
  \includegraphics[width=.45\linewidth]{beta_fromQ2-int-reg.eps}
  \caption{Value of the deformation, $\beta$, extracted from the value of the quadrupole shape invariants. In (a) we plot the value for the first three $0^+$ states. In (b) we plot the same but for the first regular and the first intruder $0^+$ state.}
  \label{fig-beta-fromQ2}
\end{figure}

Starting from the quadrupole invariant (\ref{beta-q2}), one can extract the value of the deformation $\beta$ (see, e.g., references \cite{Srebrny06,clement07,Wrzo12}).
\begin{equation}
\beta=\frac{4\, \pi\, \sqrt{q_2}}{3\, Z\, e\, r_0^2\, A^{2/3}},
\label{beta-q2}
\end{equation}
where $e$ is the proton charge and $r_0=1.2$ fm. Thus, we extract values for $\beta$ corresponding to the ground-state, $0_1^+$, and to  $0_2^+$ and $0_3^+$ states.

The resulting $\beta$ values for the first three $0^+$ states, extracted from (\ref{beta-q2}), are shown in Fig.~\ref{fig-beta-fromQ2}a. In this panel one notices how the value of $\beta$ for the ground state presents a large peak at $A=104$, whereas for the $0_2^+$ state, the behavior is rather erratic, while for $0_3^+$  the resulting value remains rather constant. A more appropriate way to present the value of $\beta$ is to select the first $0^+$ state which belongs to the regular sector, $0^+_{1reg}$, and the first $0^+$ state which belongs to the intruder sector, $0^+_{1int}$, according to the value of last column of Table \ref{tab-q-invariant}. This is presented in panel (b) of Fig.~\ref{fig-beta-fromQ2}. It is obvious that the intruder state is always more deformed than the regular one, as expected. Moreover, both exhibit a maximum at $A=104$, although the curve for the regular state is much flatter than that for the intruder one. Indeed, the curve corresponding to the regular state can be understood as having a rather constant value over the whole isotope chain, except around mid-shell. On the other hand, the curve corresponding to the intruder state shows a more abrupt increase when approaching mid-shell. Besides, there is also a sudden increase of deformation in passing from $A=94$ to $A=96$ and onwards.
\begin{figure}[hbt]
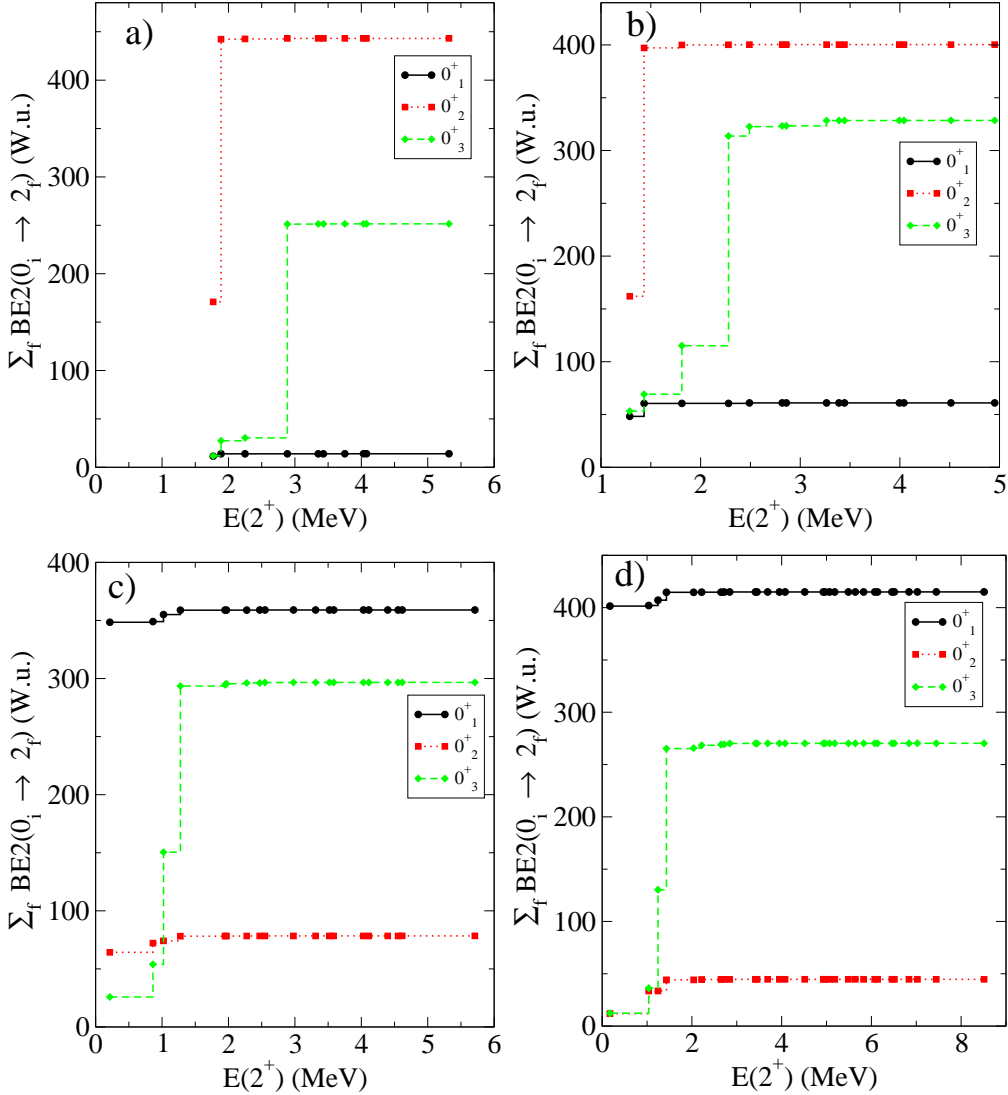

  \begin{tabular}{cc}
\includegraphics[width=.40\linewidth]{be2-strength-96ZR.eps}&\includegraphics[width=.40\linewidth]{be2-strength-98ZR.eps}\\
\includegraphics[width=.40\linewidth]{be2-strength-100ZR.eps}&\includegraphics[width=.40\linewidth]{be2-strength-108ZR.eps}
 \end{tabular}
\caption{Theoretical $\sum_f B(E2;0^+_i \rightarrow 2^+_f)$ for $i=1,2,3$ up to the $2^+$ state corresponding to the energy (theoretical) plotted in the $x$ axis for (a) $^{96}$Zr, (b) $^{98}$Zr, (c) $^{100}$Zr, and (d) $^{108}$Zr.}
  \label{fig-be2strength}
\end{figure}

\subsection{$E2$ strength function for the Zr isotopes}
\label{sec-E2strength}
In the subsection before, we have concentrated on the value of $q^2$ (\ref{q_invariant2b}) which, according to Eq.\ (\ref{q_invariant1}), corresponds to the full summed strength, connecting a given $0_i^+$ state with all possible $2_f^+$ states, thereby missing information on the different steps, $B(E2;0^+_i \rightarrow 2^+_f)$, that appear in building up the invariant,
\begin{equation}
  q_{2,i}=\sum_f B(E2;0^+_i \rightarrow 2^+_f)
  \label{be2strength}
\end{equation}
where the sum is extended over the whole Hilbert space. In Fig.\ \ref{fig-be2strength}, the partial sum of the transition strength (\ref{be2strength}) up to the $2^+$ state with a given energy, as a function of this energy, is depicted for the nuclei $^{96}$Zr, $^{98}$Zr, $^{100}$Zr, and $^{108}$Zr in panels (a), (b), (c), and (d), respectively, for the states $0_1^+$, $0_2^+$ and $0_3^+$. The number of $2^+$ states available in each nucleus is $10$, $14$, $19$, and $30$ for  $^{96}$Zr, $^{98}$Zr, $^{100}$Zr, and $^{108}$Zr, respectively. The first three cases correspond to the transition region, while the latest one, to the well deformed situation. These strength functions give a more detailed insight in the way the low-lying $0^+_i$ and $2^+_f$ states are connected and thus give hints on the character of $E2$ excitation pattern that may help to point out differences between a more ``spherical'' versus a more strongly deformed character. See also Ref.\ \cite{Schm17}  for a similar detailed study in the case of the Cd nuclei.

In  $^{96}$Zr, we notice an almost immediate reaching of the total $E2$ strength connected with the $0^+_1$ ground stated with $\sim 15$ W.u, by the $E2$ transition into the first excited $2^+_1$ state. Moving to  $^{98}$Zr, the situation remains similar, but this time, with a single step moving up to $\sim 50-60$ W.u.  For the heavier Zr nuclei, $^{100}$Zr and  $^{108}$Zr, again, there is a quick convergence, this time, however, towards a much larger values of the order of $350$ W.u. giving a clear-cut indication of a major change in the structure of the lowest-lying states as compared to the mass region situated before mass number $A=98$.

Inspecting the results in which we concentrate on, the $E2$ strength connected with the 0$^+_2$ first excited state, we notice a similar structure as is the case of the $0^+$  ground state, however, this time converging to the much larger value of $\sim 350-400$  W.u. In both $^{96}$Zr and $^{98}$Zr nuclei, a single major contribution coming from the E2 strength connects with the 2$^+_2$ excited state. For the heavier nuclei on the other hand, the convergence ends in a much smaller value, being quickly reached after just a few steps pointing out towards a major change in character between the 0$^+_1$ and  0$^+_2$ and the corresponding $2^+_1$ and $2^+_2$ states, suggesting the presence of a shape change transition.

Discussing the observed results on the E2 strength connected with the 0$^+_3$ excited state, the changing E2 strength function moving from the lighter Zr isotopes (up to mass $A=98$), is converging to almost the same accumulated value of $\approx 300$ W.u., in which the full strength remains largely concentrated in one very strong $E2$ contribution, associated with the transition into the $2^+_{4}$ state at mass $A=96$, partly shared with a non-negligible E2 contribution into the $2^+_{3}$ state. Moving up into the Zr isotopes from $A=100$ and onwards, the E2 strength is originated largely from transitions into the 2$^+_{3,4}$ states. 

We have presented these $E2$ strength function because the building up of the total $E2$ strength into the quadrupole invariant, is giving most interesting information on how the quadrupole deformation is building up in a specific way, thus pointing out the possibility to extract information on a similar ``intrinsic'' character.  

As a bottom line, we notice the almost ``dramatic'' switch in the accumulated E2 strength between the 0$^+_{1}$ and 0$^+_{2}$ states when passing beyond mass number $A=98$, with the $0^+_{3}$ not showing any major structural change. In the latter, one notices though an overall increase in the summed E2 strength, and thus in the deformation associated with the third $0^+$ state. 

\section{Conclusions and outlook}
\label{sec-conclu}
In the present paper, we have addressed the Zr isotopes, spanning a most interesting series from the $^{94}$Zr nucleus up to $^{110}$Zr. These nuclei, in due time, starting with the early experimental characterization from fission products, making use of a most extensive set of experimental techniques, allowing to characterize not only the energy spectra, but also obtaining extensive data on the electromagnetic decay properties using both (n,n’$\gamma$) lifetime measurements, recoil distance techniques, DSA, inelastic electron scattering, Coulomb excitation, mass measurements and measurement of the charge radii, at various facilities, have been studied.  Consequently, in this region a change in structure is dramatically realized, with a rather sudden evolution from more spherical properties into a region where deformation in the ground-state structure has become obvious, as indicated through the systematics of two-neutron separation energy, charge radii, as well as, the very clear $\rho^2(E0)$ variation in passing from typical values of the order of $5-10$ miliunits up to  $^{100}$Zr with one of the strongest observed $E0$ transitions with $\sim 100$ miliunits. 

Consequently, this mass region has attracted major efforts from the theoretical side, trying to understand this transition from spherical into a deformed shapes,  spanning by now more than four decades, but also theoretically, starting from early shell-model calculations, over mean-field calculations, and with the advent of large-scale shell-model calculations up to the Monte Carlo shell-model approach more recently.  In the present paper, we have studied the possibility to understand the varying nuclear characteristics within a frame in which the shell-model characteristics can be modeled using the interacting boson model, including the fact that proton excitations across the $Z=40$ subshell closure allows for a steady change from mainly vibrational-like characteristics into a region of deformation. This approach is based on the interpretation of the appearance of intruder states as a portal to the onset of deformation, as illustrated in other mass regions.  We have put the emphasis on the fact that the observed results depend on the balance between, on one side, the energy cost to redistribute the shell-model orbital occupation through np-nh excitations across the closed shell configuration and, on the other side, the quadrupole energy gain mainly originated in the proton-neutron interaction energy.

Besides a critical description of the present IBM calculations, including 2p-2h excitations and thus resulting in an enlarged model space, called the IBM-CM, a detailed comparison of energy spectra, $B(E2)$ values, with an in-depth discussion of the characteristics of the resulting wave functions as being built up from the interplay between the regular an the intruder components, to give insight in the changing nuclear properties with increasing mass number, was carried out. This is amply discussed in the present paper, in particular for the low-lying states for each of the $J$ values.  We also illustrate in detail how the unperturbed two-block system, gives rise to the fully interacting system.  Moreover, this interplay is presented in a set of plots in which a particular decomposition of the wave functions, using the U(5) basis, thereby illustrating the change in the characteristics of the members of the yrast and next higher band, is given. We have given attention to the $\rho^2(E0)$ values moving from $A=94$ up to $A=100$, concerning the most important $E0$ transitions. 

We have, in particular, given attention to the structure of the evolving ground-state structure, and its comparison with the dramatic data set on charge radii and two-neutron separation energies which are both very well described, supporting that the essential physics is included within the IBM-CM approach, as presented in the present paper.  

In view of existing mean-field calculations, we have also extracted the mean-field representation of the IBM-CM approach, making use of the intrinsic state formalism. Our results indicate a clear transition from a spherical minimum in the case of $^{98}$Zr, changing in a coexisting situation for $^{100}$Zr with a spherical and prolate deformed minimum, resulting in a single prolate deformed shape moving towards $^{102}$Zr which remains for the heavier $A=104-110$ Zr isotopes.     

Having carried out this detailed study formulated within the laboratory frame, the standard question comes up on what we can say about the ``intrinsic'' characteristics the present wave functions give rise to.  In that respect, we have calculated the quadrupole quadratic and cubic shape invariants, which result into information on both the intrinsic quadrupole moment, characterized by the value of $\beta$ and $\gamma$ which are illustrated in the present study for the lowest-lying three 0$^+$ states as well as the regular and intruder 0$^+$ state, respectively.  There is more information that we extracted besides the full invariant, namely, we have calculated the $E2$ strength function starting from the lowest three  $0^+$ states highlighting not only the convergence pattern, but, giving extra insight on how the $E2$ operator is connected to the large set of higher-lying $2^+$ states.  The result to be taken from the strength function indicates that for those cases where almost full strength is reached in one step, one can properly claim that a single intrinsic state  is ``hidden'' behind such a result.

As a conclusion (of the conclusion), we feel it rewarding that various theoretical approaches, starting from from the large-scale shell-model side with LSSM and MCSM,  various mean-field calculations as well as the present IBM-CM calculation, all seem to end up with very much the same physics results.

At present, even though already many experimental methods have been used to extract as many complementary information on the properties of the Zr isotopic series, extending by now well into a region of deformation, data that characterize the precise structure of the nuclear wave functions, such as transfer reactions, are of major interest to map out the microscopic content that has been described using a large spread of theoretical approaches.  Also, multiple Coulomb excitation allows to disentangle not just quadrupole collectivity, but serves as a fine measuring apparatus by extracting even the sign of the specific E2 matrix elements.
We emphasize the fact that both one-nucleon as well as two-nucleon transfer, in particular the (2n, $\alpha$) transfer reactions, appear as mandatory to extract information on both the single-particle energy centroids, as well as on the pairing characteristics, respectively, which are of major importance to understand the interplay of the various correlations that are active in the Zr isotopic series, indicating a transition from spherical towards deformed nuclei in both the observed energy spectra and many other observables.  Only in this way can a deeper understanding of the changing underlying structure be reached.
On the theoretical side, the intimate relations that connects the various approaches should be uncovered. We suggest that the importance of symmetries cannot be emphasized enough in order to see order in the already now existing data basis.

In the process of finishing this work, we became aware of the publication of another IBM-CM calculation by \citet{Gavr19} for the even-even Zr isotopes, where the authors have obtained very much the same results as presented here [on energy spectra, $B(E2; 2^+_1\rightarrow 0^+_1)$, isotopic shifts, and two-neutron separation energies], although they claim that the heavier Zr isotopes own an almost  O(6) symmetry, which is differing with the calculations discussed in the present work. The reason for considering $^{106-110}$Zr close to the O(6) symmetry in \cite{Gavr19} is mainly based on the presence of a $2^+$ states near to the energy of the $4_1^+$ state, in spite of a experimental value of $E(4_1^+)/E(2_1^+)$ close to $3$, which is more in favour of the start of a rotational spectrum. New experimental information in this area will  definitively be necessary to disentangle the nature of the heavier Zr isotopes.

Recent experimental studies \cite{Urba19a,Urba19b} have indicated the
possibility of gamma soft and/or even triaxial deformation showing up
in the case of $^{100}$Zr by inspecting the decay of its first excited
$0^+$ state.  If confirmed, this could be a sign of a transition from
a spherical structure into strongly deformed prolate energy spectra
beyond mass number $A=100$.  Theoretical studies showing, e.g., the
total energy surfaces such as in the case of Ref.\ \cite{Togashi16}
have shown that a QPT might appear for neutron number around
$N=70$. The latter being highlighted by a valley connecting the oblate
and prolate sides, exhibiting a shift into a triaxial shape in
$^{110}$Zr.  The paper that we already mentioned \cite{Gavr19} also
addressed the topic of triaxial shapes and illustrated this by showing
the total energy surfaces, exhibiting a tendency to evolve from a
prolate shape (from $A=100$ onwards till $106$) to move towards the
gamma direction in $^{110}$Zr, although these results are not
conclusive.

\section{Acknowledgment}
We are very grateful to A.\ Leviatan and J.L.\ Wood for enlighten discussions. This work was supported (KH) by the InterUniversity Attraction Poles Program of the Belgian State-Federal Office for Scientific, Technical and Cultural Affairs (IAP Grant  P7/12) and it has also been partially supported (JEGR) by the Consejer\'{\i}a de Econom\'{\i}a, Conocimiento, Empresas y Universidad de la Junta de Andaluc\'{\i}a (Spain) under Group FQM-370, by Centro de Estudios Avanzados en F\'isica, Matem\'atica y Computaci\'on of the  University of Huelva, and by Instituto Carlos I de F\'{\i}sica Te\'orica y Computaci\'on of the University of Granada.   

\bibliography{references}
\end{document}